\documentclass[a4paper,aps,pra,floatfix,tightenlines,superscriptaddress,twocolumn]{revtex4}

\usepackage[utf8x]{inputenc}
\usepackage{graphicx}
\usepackage{amsmath}
\usepackage{amsthm}
\usepackage{amsfonts}
\usepackage{amssymb}
\usepackage{paralist}
\usepackage{bbm}
\usepackage{latexsym}
\usepackage{multirow}
\usepackage[bookmarks = true, colorlinks = true, urlcolor=blue]{hyperref}



\begin{document}

\title{Cavity cooling of an optically levitated nanoparticle}

\author{Nikolai Kiesel}
\thanks{These authors contributed equally to this work.}
\affiliation{Vienna Center for Quantum Science and Technology (VCQ), Faculty of
Physics, University of Vienna, Boltzmanngasse 5, A-1090 Vienna, Austria}
\author{Florian Blaser}
\thanks{These authors contributed equally to this work.}
\affiliation{Vienna Center for Quantum Science and Technology (VCQ), Faculty of
Physics, University of Vienna, Boltzmanngasse 5, A-1090 Vienna, Austria}
\author{Uro\v s Deli\' c}
\affiliation{Vienna Center for Quantum Science and Technology (VCQ), Faculty of
Physics, University of Vienna, Boltzmanngasse 5, A-1090 Vienna, Austria}
\author{David Grass}
\affiliation{Vienna Center for Quantum Science and Technology (VCQ), Faculty of
Physics, University of Vienna, Boltzmanngasse 5, A-1090 Vienna, Austria}
\author{Rainer Kaltenbaek}
\affiliation{Vienna Center for Quantum Science and Technology (VCQ), Faculty of
Physics, University of Vienna, Boltzmanngasse 5, A-1090 Vienna, Austria}
\author{Markus Aspelmeyer}
\affiliation{Vienna Center for Quantum Science and Technology (VCQ), Faculty of
Physics, University of Vienna, Boltzmanngasse 5, A-1090 Vienna, Austria}
\maketitle

\textbf{The ability to trap and to manipulate individual atoms is
at the heart of current implementations of quantum simulations \cite{Blatt2012,Bloch2012},
quantum computing \cite{HAFFNER2008,Kielpinski2002}, and long-distance
quantum communication \cite{Kimble2008,Ritter2012,Stute2013,Hofmann2012}.
Controlling the motion of larger particles opens up yet new avenues
for quantum science, both for the study of fundamental quantum phenomena
in the context of matter wave interference \cite{Gerlich2011,Hornberger2012},
and for new sensing and transduction applications in the context of
quantum optomechanics \cite{Aspelmeyer2012,Aspelmeyer2013}. Specifically,
it has been suggested that cavity cooling of a single nanoparticle
in high vacuum allows for the generation of quantum states of motion
in a room-temperature environment \cite{Chang2010,Romero-Isart2010,Romero-Isart2011a}
as well as for unprecedented force sensitivity \cite{Geraci2010,Arvanitaki2013}.
Here, we take the first steps into this regime. We demonstrate cavity
cooling of an optically levitated nanoparticle consisting of approximately
$10^{9}$ atoms. The particle is trapped at modest vacuum levels of
a few millibar in the standing-wave field of an optical cavity and is
cooled through coherent scattering into the modes of the same cavity
\cite{Horak1997a,Vuletic2000a}. We estimate that our cooling rates
are sufficient for ground-state cooling, provided that optical trapping
at a vacuum level of $10^{-7}$ millibar can be realized in the future,
e.g., by employing additional active-feedback schemes to stabilize
the optical trap in three dimensions \cite{Ashkin77,Li2011b,Gieseler2012,Koch2010}.
This paves the way for a new light-matter interface enabling room-temperature
quantum experiments with mesoscopic mechanical systems.}

Cooling and coherent control of single atoms inside an optical cavity
are well-established techniques within atomic quantum optics \cite{Ye1999,McKeever2003,Maunz2004a,Leibrandt2009,Stute2012}.
The main idea of cavity cooling relies on the fact that the presence
of an optical cavity can resonantly enhance scattering processes of
laser light that deplete the kinetic energy of the atom, specifically
those processes where a photon that is scattered from the atom is
Doppler-shifted to a higher frequency. It was realized early on that
such cavity-enhanced scattering processes can be used to achieve laser
cooling even of objects without exploitable internal level structure
such as molecules and nanoparticles {\cite{Horak1997a,Hechenblaikner1998,Gangl2000,Vuletic2000a}.
For nanoscale objects, cavity cooling has been demonstrated in a series
of recent experiments with nanobeams \cite{Favero2009b,Anetsberger2009a,Chan2011b}
and membranes of nm-scale thickness (e.g.~\cite{Thompson_etal2008,Teufel2011}).
To guarantee long interaction times with the cavity field these objects
were mechanically clamped, which however introduces additional dissipation
and heating through the mechanical support structure. As one consequence,
quantum signatures have thus far only been observed in a cryogenic
environment \cite{Purdy2013,Safavi-Naeini2012}. Freely suspended
particles can circumvent this limitation and allow for far better
decoupling of the mesoscopic object from the environment. This has
been successfully implemented for atoms driven at optical frequencies
far detuned from the atomic resonances, both for the case of optically
trapped single atoms \cite{Leibrandt2009,Maunz2004a} and for clouds
of up to $10^{5}$ ultracold atoms \cite{Murch2008,Purdy2010,Schleier-smith2011a}.
In contrast to such clouds, massive solid objects provide access to
a new parameter regime: on the one hand, the rigidity of the object
allows to manipulate the center-of-mass motion of the whole system,
thus enabling macroscopically distinct superposition states \cite{Romero-Isart2010,Romero-Isart2011a,Kaltenbaek2012a};
on the other hand the large mass density of solids concentrates many
atoms in a small volume of space, which provides new perspectives
for force sensing \cite{Arvanitaki2013,Geraci2010}. In our work,
we have now extended the scheme to dielectric nanoparticles comprising
up to $10^{9}$ atoms. By using a high-finesse optical cavity for
both optical trapping and manipulation we demonstrate, for the first
time, cavity-optomechanical control, including cooling, of the center-of-mass
(CM) motion of a levitated solid object without internal level structure.

\begin{figure*}
\includegraphics[width=1.0\linewidth]{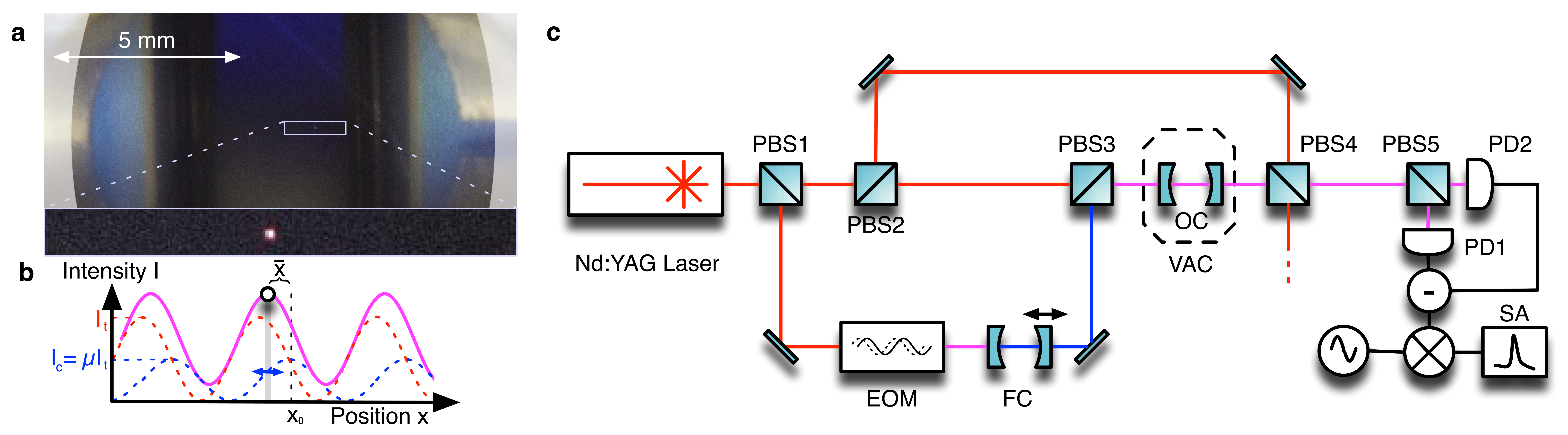}

\caption{\textbf{Optical trapping and readout of a nanoparticle in a Fabry-Perot
cavity. (a) Nanoparticle in a cavity.} A photo of our near-confocal
Fabry-Perot optical cavity (OC) (F=76000; $L=\frac{c}{2\mathrm{FSR}}=$10.97
mm, determined via the free spectral range FSR). The white-shaded
areas indicate the curvature of the cavity mirrors. The optical field
between the mirrors traps a nanoparticle. The enlarged inset shows
light scattered by the nanoparticle. \textbf{(b) Schematics of two-mode optical
trap and dispersive coupling.} Two optical fields form standing-wave
intensity distributions along the optical cavity axis (dashed lines;
blue: control beam; red: trapping beam). Because of their different
frequencies, the intensity maxima of the two fields are displaced
with respect to each other. A nanoparticle is trapped at the maximum
of the total intensity distribution (purple solid line). Since the
trapping beam is more intense than the control beam, the nanoparticle
is trapped at a distance $\overline{x}\neq0$ away from the control-beam intensity
maximum $x_{0}$. As a consequence, the nanoparticle oscillates within
a region where the control-beam intensity varies with the particle
position (blue arrow), resulting in linear dispersive coupling (see
main text and appendix). The displacement $\overline{x}$ depends on the
ratio between the intensity maxima of the two fields \textbf{(c) Experimental
setup.} A Nd:YAG laser ($\lambda=1064$~nm) is split into three
beams at the polarizing beam splitters PBS1 and PBS2 (for simplicity
waveplates not shown in the figure). The transmitted beam is used
to lock the laser to the TEM00 mode of the OC and provides the trapping
field for the nanoparticle. The beam reflected at PBS1 is used to
prepare the control beam, which is frequency-shifted by $\delta\omega$
close to the adjacent cavity resonance of the TEM00 mode, i.e., $\textrm{\ensuremath{\delta\omega}=FSR}+\Delta$
($\Delta$: detuning from cavity resonance). The single-frequency
sideband at $\delta\omega$ is created using an electro-optical modulator
(EOM) followed by optical amplification in fiber (not shown) and transmission
through a filtering cavity (FC) with an FWHM linewidth of $2\text{\ensuremath{\pi\times}}500$~MHz.
The control and trapping beams are overlapped at PBS3 and transmitted
through the OC with orthogonal polarizations. The OC is mounted inside
a vacuum chamber (VAC). When a nanoparticle is trapped in the optical
field in the cavity, its center-of-mass (CM) motion introduces a phase
modulation on the control beam. To detect this signal, we perform
interferometric phase readout of the control beam: At PBS4 the trapping
beam is separated from the control beam and overlapped with the local
oscillator (LO). After rotating the polarization, the control beam
and the LO are mixed at PBS 5. High-frequency InGaAs photo detectors
PD1 and PD2 detect the light in both output ports of PBS5. We mix
the difference signal of the two detectors with an electronic local
oscillator of frequency $\mathrm{FSR}+\Delta$ and record the noise
power spectrum of resulting signal at a spectrum analyzer (SA) (see
Methods).}

\label{Fig1} 
\end{figure*}

To understand the principle of our approach, consider a dielectric
spherical particle of radius $r$ smaller than the optical wavelength
$\lambda$. Its finite polarizability $\xi=4\pi\epsilon_{0}r^{3}Re\left\{ \frac{\epsilon-1}{\epsilon+2}\right\} $($\epsilon$:
dielectric constant; $\epsilon_{0}$: vacuum permitivity) results
in an optical gradient force that allows to trap particles in the
intensity maximum of an optical field \cite{Ashkin2007}. The spatial
modes of an optical cavity provide a standing-wave intensity distribution
along the cavity axis $x$. A nanoparticle that enters the cavity
will be pulled towards one of the intensity maxima, located a distance
$x_{0}$ from the cavity center. For the case of a Gaussian (TEM00)
cavity mode, the spatial profile will result in radial trapping around
the cavity axis, hence providing a full 3D particle confinement. In
addition, Rayleigh scattering off the particle into the cavity mode
induces a dispersive change in optical path length and shifts the
cavity resonance frequency by $U_{0}(x_{0})=\frac{\omega_{cav}^{\textrm{}}\xi}{2\epsilon_{0}V_{cav}}\big(1+\frac{x_{0}^{2}}{x_{R}^{2}}\big)$
\cite{Nimmrichter2010b}($\omega_{cav}$: cavity frequency; $V_{cav}$:
cavity mode volume; $x_{R}$: cavity-mode Rayleigh length). This provides
the underlying optomechanical coupling mechanism between the CM motion
of a particle moving along the cavity axis and the photons of a Gaussian
cavity mode. The resulting interaction Hamiltonian is 
\begin{eqnarray*}
H_{int} & =-\hbar U_{0}(x_{0})\hat{n}\textrm{sin}^{2}(kx_{0}+k\overline{x}+k\hat{x}),
\end{eqnarray*}
where we have allowed for a mean displacement $\overline{x}$ of the
nanoparticle with respect to the intensity maximum $x_{0}$ ($\hat{x}$:
CM position operator of the trapped nanoparticle; $k$=$\frac{2\pi}{\lambda}$:
wavenumber of the cavity light field; $\hat{n}$: cavity photon number
operator). For the case of a single optical cavity mode, the particle
is trapped at an intensity maximum ($\overline{x}=0$) and, for small
displacements, only coupling terms that are quadratic in $\hat{x}$
are relevant \cite{Thompson_etal2008}. Linear coupling provides intrinsically
larger coupling rates and can be exploited for various quantum control
protocols \cite{Romero-Isart2011}. However, it requires to position
the particle outside the intensity maximum of the field. This can
be achieved for example by an optical tweezer external to the cavity
\cite{Romero-Isart2010}, by harnessing gravity in a vertically mounted
cavity \cite{Barker2010a} or by using a second cavity mode with longitudinally
shifted intensity maxima \cite{Chang2010,Romero-Isart2010}. 

\begin{figure}
\includegraphics[width=1.0\linewidth]{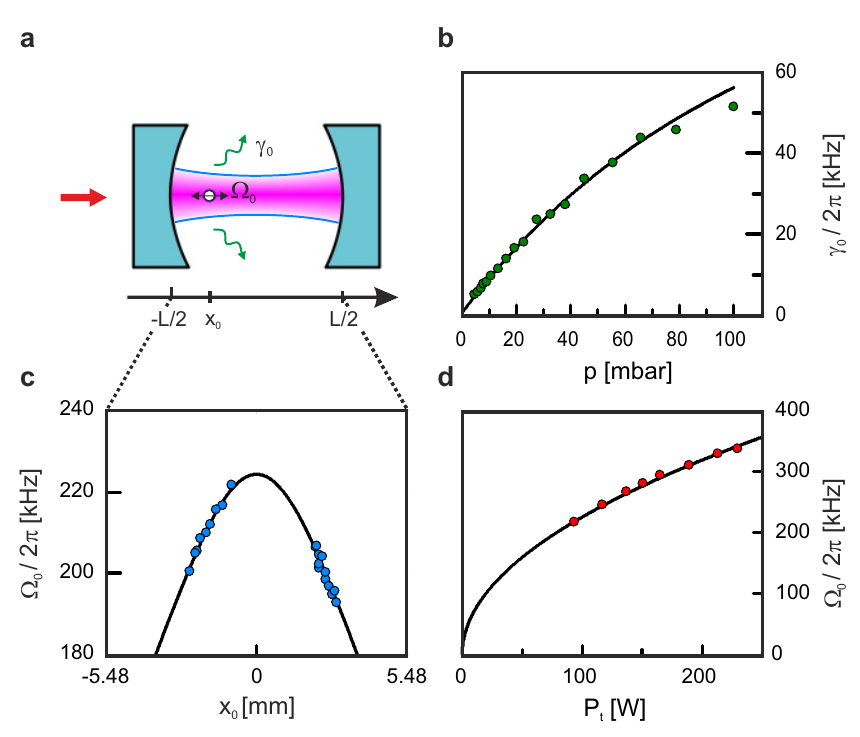}
\caption{\textbf{Experimental characterization of the nanoparticle cavity trap
. (a) Schematic of the trap configuration. }An optical cavity of length
$L=10.97$ mm is driven on resonance of a Gaussian TEM00 cavity mode
by a laser with a wavelength of $\lambda=1064$ nm. The nanoparticle
is optically trapped at position $x_{0}$. Its center-of-mass motion
in the axial direction of the cavity is described by a harmonic oscillator
with a frequency $\Omega_{0}$ and an amplitude of approximately 10
nm . In addition, the nanoparticle experiences collisions with the
surrounding gas resulting in a damping rate $\gamma_{0}$. \textbf{(b)
Mechanical damping }$\gamma_{0}$\textbf{ as a function of pressure}.
The solid line is a fit of kinetic gas theory to the data (see appendix \ref{app:gas}).\textbf{ (c)} \textbf{Position-dependent trapping frequency.
}The waist of the optical mode expands from approximately $41\mu m$
at the cavity center to $61\mu m$ at the cavity mirrors, resulting
in a position-dependent trapping potential. Here, we show the corrsponding
change of the trapping frequency $\Omega_{0}$ with the position of
the nanoparticle. \textbf{(d)} \textbf{Power-dependent trapping frequency.}
We experimentally show the dependence of the trapping frequency on the intracavity power
$P_{\textrm{t}}$. The solid lines in Fig.~\textbf{c, d} are based
on the theoretical model as described in the main text, with a scaling
factor as the only free fit parameter.}
\label{Figure 2}
\end{figure}

We follow the latter approach and operate the optical cavity with
two longitudinal Gaussian modes of different frequency, namely, a
strong ``trapping field'' to realize a well-localized optical trap at
one of its intensity maxima, and a weaker ``control field'' that couples
to the particle at a shifted position $\overline{x}\ne0$. For localization
in the Lamb-Dicke regime ($k^{2}\langle\hat{x}^{2}\rangle\ll1$) this
yields \cite{Stamper-Kurn2012,Aspelmeyer2013} linear optomechanical
coupling between the trapped particle and the control field at a rate
$g_{0}=U_{0}(x_{0})\sin(2k\overline{x})k\sqrt{\frac{\hbar}{m\Omega_{0}}}$
per photon ($m$: nanoparticle mass; $\Omega_{0}$: frequency of CM
motion). Detuning of the control field from the cavity resonance by
a frequency $\Delta=\omega_{cav}^{\textrm{}}-\omega_{\textrm{c}}$($\omega_{c}:$
control field frequency) results in the well-known dynamics of cavity
optomechanics \cite{Aspelmeyer2013}. Specifically, the position dependence
of the gradient force will change the stiffness of the optical trap,
shifting $\Omega_{0}$ to an effective frequency $\Omega_{\textrm{eff}}$
(optical spring), and the cavity-induced retardation of the force
will introduce additional optomechanical (positive or negative) damping
on the particle motion. From a quantum-optics viewpoint, the oscillating
nanoparticle scatters photons into optical sidebands of frequencies
$\omega_{\textrm{c}}\pm\Omega_{0}$ at rates $A_{\pm}=\frac{1}{4}\frac{g_{0}^{2}\langle\hat{n}\rangle\kappa}{(\kappa/2)^{2}+(\Delta\pm\text{\ensuremath{\Omega_{0}}})^{2}}$,
known as Stokes and anti-Stokes scattering, respectively ($\kappa$:
FWHM cavity linewidth). For $\Delta>0$ (red detuning) anti-Stokes
scattering becomes resonantly enhanced by the cavity, effectively
depleting the kinetic energy of the nanoparticle motion via a net
laser-cooling rate of $\Gamma=A_{-}-A_{+}$. In the following, we
demonstrate all these effects experimentally with an optically trapped
silica nanoparticle. 

\begin{figure*}[th]
\includegraphics[width=1.0\linewidth]{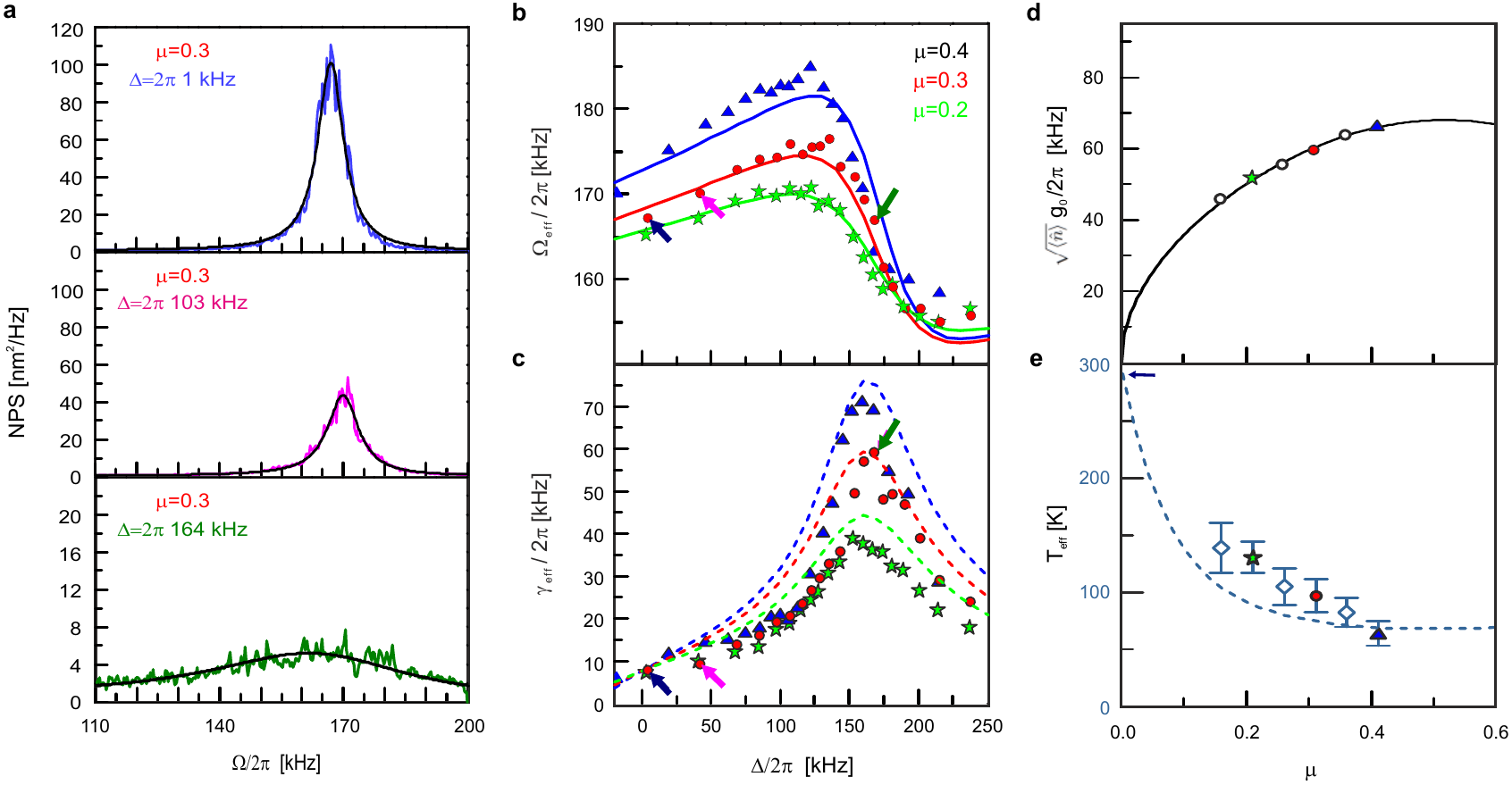}
\caption{\textbf{Cavity-optomechanical control and cooling of a nanoparticle.}
We obtain noise power spectra (NPS, see panel
(a)) of the nanoparticle's center-of-mass motion for different settings of the control-beam
power $P_{c}$ and detuning $\Delta$. During each measurement, $\mu=\frac{P_{c}}{P_{t}}$ was kept constant (
$P_{t}$: trapping beam power). Based
on these NPS, we determine the effective mechanical frequency $\Omega_{\text{eff}}$ and linewidth $\gamma_{\text{eff}}$ of the optomechanical system, and its effective temperature $T_{\textrm{eff}}$. We study
the modification of these spectra caused by optomechanical interaction
in panels (b), (c) and (e). Based on the data in panel (b) we infer
the power-dependent strength of optomechanical coupling in panel (d).\textbf{
(a) Mechanical noise power spectra.} Shown are examples
of the mechanical NPS measured for constant control-beam power ($\mu=0.3$)
at three different detunings $\Delta$ with respect to the cavity
resonance frequency. The detuning results in a significant modification
of the NPS due to optomechanical effects. Note that scale is changed by a factor of 5 in the bottom plot in panel (a). In order to determine the
effective mechanical frequency $\Omega_{\text{eff}}$ and linewidth
$\gamma_{\text{eff}}$ of the optomechanical system, we fit the NPS
of an harmonic oscillator (black
solid lines) to this data. We infer the value of the effective temperature $T_{\textrm{eff}}$
from the equipartition theorem via direct integration of the NPS (see appendix \ref{app:evaluation}).\textbf{ (b) Optical spring.} When
the control beam is red-detuned from the cavity resonance ($\Delta>0$),
we observe a characteristic modification of the mechanical frequency
$\Omega_{\text{eff}}$. The solid lines in (b) correspond to a theoretical
model that is fitted to the data for each value of $\mu$. The optomechanical coupling $g_{\text{0}}\sqrt{\langle\hat{n}\rangle}$
is one of the fit parameters (see appendix \ref{app:evaluation}). Based
on these results for the optical spring, we calculate the theoretical
expectations for $\gamma_{\text{eff}}$ and $T_{\textrm{eff}}$, which
are shown as dashed lines in panels (c) and (e). \textbf{(c) Optomechanical
damping}. Linewidth broadening of the mechanical resonance as a function
of the detuning $\Delta$. \textbf{(d) Optomechanical coupling. }We
infer the optomechanical coupling rate $g_{\text{0}}\sqrt{\langle\hat{n}\rangle}$
from the strength of the optical spring (panel (b)) and show its dependence
on the power ratio $\mu$. This relation depends on the position $x_{0}$
of the nanoparticle in the cavity. For the data presented here, we
determine $x_{0}=1.56\pm0.14$ mm (see appendix \ref{app:position}).
We find very good agreement between the data and the theoretical model,
where only the nanoparticle polarizability serves as a fit parameter
(solid line; also see appendix \ref{app:evaluation}).
\textbf{(e) Cavity cooling. }The decrease in effective temperature
$T_{\textrm{eff}}$ is shown for increasing control-beam power. To
obtain a good estimate of the measurement error, we average over measurements
taken for detunings between $\Delta=100-150$ kHz (see appendix \ref{app:evaluation}). The dashed line is a theoretical prediction based on
the parameters obtained from the fit to the optical spring data (panel
(b)).}

\label{Fig3} 
\end{figure*}

As is shown in Figure~\ref{Fig1}, our setup comprises a high-finesse
Fabry-Perot cavity (Finesse $F=76000$; $\kappa=2\pi\times180$ kHz)
that is mounted inside a vacuum chamber kept at a pressure between
1 and 5 mbar. Airborne silica nanoparticles (specified with radius
$r=127\pm13$ nm) are emitted from an isopropanol solution via an
ultrasonic nebulizer and are trapped inside the cavity in the standing
wave of the trapping field (see Methods Section). To achieve the desired
displacement between the intensity maxima of trapping field and control
field ($\overline{x}\neq0$), we use the adjacent longitudinal cavity
mode for the control beam, i.e. the cavity mode shifted by approximately
one free spectral range $\textrm{FSR}=\frac{c}{2L}\approx13.67$ GHz
in frequency from the trapping beam ($c$: vacuum speed of light;
$L$: cavity length). Depending on the distance from the cavity center
$x_{0}$, the two standing-wave intensity distributions are then shifted
with respect to each other by $\frac{\lambda}{2L}(x_{0}+L/2)$ (Figure~\ref{Fig1}c). For example, to achieve maximal coupling $g_{0}$ for
weak control beam powers, i.e. for $\mu=\frac{P_{c}}{P_{t}}\ll1$
($P_{\textrm{c(t)}}$: Power of control (trapping) beam in the cavity),
the nanoparticle needs to be positioned at $x_{0}=L/4$, where the
antinodes of the two beams are separated by $\lambda/8$ \cite{Chang2010,Romero-Isart2010}.
Note that when the control beam is strong enough to significantly
contribute to the optical trap ($\mu\gtrapprox0.1$), the displacement
$\bar{x}$ and both $\Omega_{0}$ and $g_{0}$ are modified when $\mu$
is changed \cite{Purdy2010} . The exact dependence of these optomechanical
parameters on $\mu$ depends on $x_{0}$ (see appendix \ref{app:optomech} and \cite{Pender2012, Monteiro2013}). 

The optomechanical coupling between the control field and the particle
can be used to both manipulate and detect the particle motion. Specifically,
the axial motion of the nanoparticle generates a phase modulation
of the control field, which we detect by heterodyne detection (see
methods section). We reconstruct the noise power spectrum (NPS) of
the mechanical motion by taking into account the significant filtering
effects exhibited by the cavity (arising from the fact that $\kappa\approx\Omega_{0}$)
on the transmitted control beam (\cite{Paternostro2006} and appendix \ref{app:optomech}). The inferred position sensitvity of our readout scheme
for a nanoparticle of approx. 170 nm radius is $4\textrm{ pm}/\sqrt{\text{{Hz}}}$,
which is likely limited by classical laser noise (see below). 

The properties of our optical trap are summarized in Figure \ref{Figure 2}.
The influence of the control beam on the trapping potential is purposely
kept small by choosing $\mu\approx0.1$ and $\Delta\approx0$. We
expect that the axial mechanical frequency $\Omega_{0}$ depends both
on the power of the trapping beam $P_{\textrm{t}}$ and on $x_{0}$
through the cavity beam waist $W(x_{0})$ via $\Omega_{0}=\sqrt{\frac{12k^{2}}{c\pi}Re(\frac{1}{\rho}\frac{\epsilon-1}{\epsilon+2})}\cdot\sqrt{\frac{P_{\textrm{t}}}{\pi W(x_{0})}}$
\cite{Chang2010,Romero-Isart2010}, in agreement with our data. The
damping $\gamma_{0}$ of the mechanical resonator is dominated by
the ambient pressure of the background gas down to a few millibar
(Fig.~\ref{Fig1}c). Below these pressures the nanoparticle is not
stably trapped anymore, while trapping times up to several hours can
be achieved at a pressure of a few millibar. This is a known, yet
unexplained phenomenon \cite{Li2011b,Pender2012,Gieseler2012}. Reproducible
optical trapping at lower pressure values has thus far only been reported
using feedback cooling in three dimensions for the case of nanoparticles
\cite{Li2011b,Gieseler2012} or, without feedback cooling, with particles
of at least $20\,\mu$m radius \cite{Ashkin1976a}. 

We finally demonstrate cavity-optomechanical control of our levitated
nanoparticle. All measurements have been performed with the same particle
for an intra-cavity trapping beam power $P_{t}$ of approx. $55$
W and at a pressure of $p\approx4$ mbar. This corresponds to a bare
mechanical frequency $\Omega_{0}/2\pi=165\pm3$ kHz and an intrinsic
mechanical damping rate $\gamma_{0}/2\pi=7.2\pm0.8$ kHz, respectively.
Figure \ref{Fig3}a shows the dependence of a typical noise power spectrum
(NPS) of the particle's motion upon detuning of the control field. Note
that the power ratio $\mu$ between trapping beam and control beam
is kept constant, which is achieved by adjusting the control-beam
power for different detunings. The amplitude scale, as well as the
temperature scale in Figure \ref{Fig3}e, is calibrated through the
NPS measurement performed close to zero detuning ($\Delta=1$ kHz;
blue NPS in Fig.~\ref{Fig3}a by using the equipartition theorem
for $T=293K$. This is justified by an independent measurement that
verifies thermalization of the center of mass (CM) mode at zero detuning
for our parameter regime (see appendix \ref{app:gas}). Both the
inferred effective mechanical frequency $\Omega_{\text{eff}}$ (Figure
\ref{Fig3}b and the effective mechanical damping $\gamma_{\text{eff}}$
(Figure \ref{Fig3}c show a systematic dependence on the detuning
$\Delta$ of the control beam, in good agreement with the expected
dynamical backaction effects for linear optomechanical coupling (see appendix \ref{app:optomech}). A fit of the expected theory curve to
the optical spring data allows estimating the strength of the optomechanical
coupling for different values of $\mu$ (Figure \ref{Fig3}d). If
the position $x_{0}$ of the nanoparticle in the cavity is known,
then this behaviour is uniquely determined by $U_{0}(x_{0})$. For
a particle position $x_{0}=1.56\pm0.14$ mm, which was determined
independently with a CCD camera, we find $U_{0}(x_{0})=2\pi\times(145\pm2)$
kHz. These values allow to infer a nanoparticle displacement $\overline{x}\approx0.15\times(\lambda/2)=77$
nm, yielding a fundamental single-photon coupling rate $g_{0}\approx2\pi\times1.2$
Hz (for $\mu\rightarrow0$). Assuming a (supplier specified) material
density of $\rho=1950\text{ g/cm}^{3}$and a dielectric constant $\epsilon_{\text{\ensuremath{\mathrm{SiO_{2}}}}}=2.1$,
our results indicate a single trapped nanoparticle of radius $r\approx169$
nm. 

The red-detuned driving of the cavity by the control laser also cools
the CM motion of the levitated nanoparticle through coherent
scattering into the cavity modes. Figure \ref{Fig3}e shows the resulting
effective temperature as deduced from the area of the NPS of the mechanical
motion by applying the equipartition theorem. The experimental data
is well in agreement with the expected theory for cavity cooling (see appendix \ref{app:optomech}). We achieve cooling rates of up to $\Gamma=2\pi\times49$
kHz and effective optomechanical coupling rates of up to $g_{0}\sqrt{\langle\hat{n}_{c}\rangle}=2\pi\times66$
kHz ($\langle\hat{n}_{c}\rangle$: mean photon number in control field),
comparable to state-of-the-art clamped mechanical systems in that
frequency range \cite{Aspelmeyer2013}. The demonstrated cooling performance,
with a minimal CM-mode temperature of $64\pm5$ K, is only limited
by damping through residual gas pressure that results in a mechanical
quality of $Q=\frac{\Omega{}_{0}}{\gamma_{0}}\approx25$. Recent experiments
\cite{Li2011b,Gieseler2012} impressively demonstrate, that lower
pressures can be achieved when cooling is applied in all three spatial
dimensions. Given the fact that our cavity-induced longitudinal cooling
rate is comparable to the feedback cooling rates achieved in those
experiments, a combined scheme should eventually be capable of performing
quantum experiments at moderately high vacuum levels. For example,
our cooling rate is in principle sufficient to obtain cooling to the
quantum ground state of the CM-motion starting from room temperature
with a longitudinal mechanical quality factor of $Q\approx10^{9}$,
i.e., a vacuum level of $10^{-7}$ mbar. Such a performance is currently
out of reach for other existing cavity optomechanical systems with
comparable frequencies. In addition, even larger cooling rates are
expected when both beams are red-detuned to cooperatively cool the
nanoparticle motion \cite{Pender2012}. 

Our experiment constitutes a first proof of concept demonstration
in that direction. We envision that once this level of performance
is achieved levitated nanoparticles in optical cavities will provide
a room-temperature quantum interface between light and matter, along
the lines proposed in \cite{Chang2010,Akram2010,Romero-Isart2010,Romero-Isart2011},
with new opportunities for macroscopic quantum experiments in a regime
of large mass \cite{Kaltenbaek2012a,Romero-Isart2011a,Romero-Isart2011b}.
The large degree of optomechanical control over levitated objects
may also enable applications in other areas of physics such as for
precision force sensing \cite{Arvanitaki2013,Geraci2010} or for studying
non-equilibrium dynamics in classical and quantum many-body systems \cite{Lechner2012}.

\section*{Acknowledgements}
We would like to thank O.~Romero-Isart, A.~C.~Pflanzer, J.~I.~Cirac, P.~Zoller, H.~Ritsch, C.~Genes, S.~Hofer, G.~D.~Cole, W.~Wieczorek, M.~Arndt, T.~Wilk for stimulating discussions and support and J.~Schmöle for his graphical contributions. We acknowledge funding from the Austrian Science Fund FWF (START, SFB FOQUS), the European Commission (IP Q-ESSENCE, ITN cQOM), the European Research Council (ERC StG QOM), the John Templeton Foundation (RQ-8251) and the European Space Agency (AO/1-6889/11/NL/CBi). N.~K.~acknowledges support by the Alexander von Humboldt Stiftung. U.~D., D.~G. acknowledge support by the FWF through the Doctoral Programme CoQuS. R.~K.~acknowledges support from the Austrian Academy of Sciences (APART) and the European Commission (Marie Curie). M.~A. and R.~K. acknowledge support through the Keck Institute for Space Studies.

\section*{Author Information}
Correspondence and requests for materials should be addressed to N.~K.~(e-mail: 
nikolai.kiesel@univie.ac.at) or M.~A.~(e-mail: markus.aspelmeyer@univie.ac.at).

\section*{Methods}

\subsection*{\textbf{Loading of nanoparticles into the optical cavity trap}}

For our experiment we use silica nanospheres (Corpuscular Inc.) with
a radius of $r=127\pm13$ nm , which are provided in an aqueous solution
with a mass concentration of 10\%. We dilute the solution with isopropanol
to a mass concentration of $10^{-7}$ and keep it for approximately
30 min in an ultrasonic bath before usage. To obtain airborne nanoparticles,
an ultrasonic medical nebulizer (Omron Micro Air) emits droplets from
the solution with approximately $3\mu\textrm{m}$ size \cite{Summers2008, Monteiro2013}. On average, the
number of nanospheres per droplet is then approximately $5\cdot10^{-4}$. 

The nanospheres are loaded into the vacuum chamber by spraying the
droplets through an inlet valve at the end of a 6mm thick, 90cm long
steel tube. We keep the pressure inside the vacuum chamber between
1 and 5 mBar via manual control of both the inlet valve connected
to the nebulizer and the outlet valve connected to the vacuum pumps.
During the loading process, the trapping laser is kept resonant with
the cavity at the desired intracavity power for optical trapping.
The low pressure minimizes pressure-induced fluctuations of the optical
path length, which significantly simplifies locking the laser to the
cavity. 

Trapping in the conservative potential of the standing-wave trap is
only possible with an additional dissipative process, which is provided
fully by damping due to the remaining background gas. Within a few
seconds after opening the valve, nanospheres get optically trapped.
The standing-wave configuration provides multiple trapping positions.
Trapped nanoparticles are detected by a CCD-camera, which is also
used to determine their position $x_{0}$ (see appendix \ref{app:position}).
If initially more than one position in the cavity is occupied, blocking
the trapping beam for short intervals allows loosing surplus particles
for our measurements. To move the trapped particle to different positions
along the cavity, we blue-detune the control laser to heat the CM
degree of freedom of the particle. The ``hot'' particle moves across
the standing wave until the control beam is switched off and the particle
stays trapped at its new position (see Figure \ref{Figure 2}b).

\subsection*{Readout of control beam}

For the position readout of the nanoparticle motion, we rely on the
dispersive interaction with the control-field cavity mode. The control
laser beam is initially prepared with a frequency difference of $\delta\omega\approx2\pi\times13.67$
GHz with respect to the original laser frequency $\omega_{0}$. When
the control beam is transmitted through the cavity, it experiences
a phase shift according to its detuning from the resonance $\omega_{cav}$.
Because the particle position in the cavity modifies the cavity resonance
frequency $\omega_{cav}$, a phase readout of the transmitted control
beam allows reconstructing the nanoparticle's motion. To detect the
phase modulation introduced by the particle motion along the cavity,
we first mix the control beam with a local oscillator (LO, 3.15 mW;
control beam power $<0.1$mW) at frequency $\omega_{0}$ at PBS5 (Fig.~\ref{Fig1}). In the output ports of PBS5, we then detect the optical
signal at photodetectors PD1 and PD2 (Discovery Semiconductor Inc.
DSC-R410), which are fast enough to process the beat signal at frequency
$\delta\omega$. Their difference signal $l(t)$, i.e., the heterodyne
measurement outcome, contains the beat signal, whose phase $\phi_{\text{opt}}$
is determined by the unknown path difference between the LO and the
control beam. The beat signal carries sidebands representing the amplitude
and phase modulation imprinted on the control beam by the optomechanical
system. We demodulate $l(t)$ with an electronic local oscillator
(ELO) with frequency $\delta\omega$ and phase $\phi_{\text{ELO}}$
(relative to the beat signal). From the resulting signal $s_{\textrm{opt}}(t)$,
we extract the phase modulation of $l(t)$ by adjusting $\phi_{\text{ELO}}$
such that the total phase $\phi_{\text{ELO}}+\phi_{\text{opt}}=\pi/2$
. This is achieved by locking the DC part of $\langle s_{\textrm{opt}}(t)\rangle$
to zero. We record the NPS of $s_{\textrm{opt}}(t)$ with a spectrum
analyzer, which allows reconstructing the NPS of the nanoparticle's
motion in post processing.

\onecolumngrid
\appendix

\section{Description of the Optomechanical System}
\label{app:optomech}

\subsection*{Optomechanical Hamiltonian}

To describe our experiment theoretically, we consider a nanoparticle
that is optically trapped within a Fabry-Perot cavity. Two laser beams
drive adjacent $\text{{TEM}}{}_{00}$ cavity modes. One beam is used
for optical trapping (trapping beam), the other for optomechanical
control and readout of the nanoparticle center-of-mass motion (control
beam). The two mode's resonance frequencies differ by one $\text{{FSR}}=\frac{c}{2L}$
($L$: cavity length). In the most general case, the two lasers can
be detuned from the respective cavity resonance frequency by $\Delta_{t}$
and $\Delta_{c}$ ($\Delta_{c(t)}$: detuning of the control (trapping)
beam). The system is described using the following Hamiltonian \cite{Monteiro2013}:
\begin{eqnarray}
\hat{H}/\hbar & = & \Delta_{t}\hat{a}_{t}^{\dagger}\hat{a}_{t}+\Delta_{c}\hat{a}_{c}^{\dagger}\hat{a}_{c}+\frac{\hat{p}_{m}^{2}}{2m\hbar}-U_{0}\hat{a}_{t}^{\dagger}\hat{a}_{t}\sin^{2}(k_{t}\hat{x})\nonumber \\
 &  & -U_{0}\hat{a}_{c}^{\dagger}\hat{a}_{c}\sin^{2}(k_{c}\hat{x})+iE_{t}(\hat{a}_{t}^{\dagger}-\hat{a}_{t})+iE_{c}(\hat{a}_{c}^{\dagger}-\hat{a}_{c}),\label{eq: Ham1}
\end{eqnarray}

where $U_{0}$ can be understood as the cavity resonance frequency
shift introduced by a nanoparticle that is located at the intensity
maximum at the center of the optical cavity. At the same time, $\hbar U_{0}$
is also the trap depth created by a single intracavity photon ($\hat{a}_{c(t)}^{\dagger}$
/ $\hat{a}_{c(t)}$: creation/annihilation operator of the control
(trapping) field in the cavity; $m$: mass of the nanoparticle; $\hat{x}$
($\hat{p}_{m}$): position (momentum) operatior of the nanoparticle's
CM; $k_{c(t)}$/$E_{c(t)}$: wavenumber/driving field of the control
(trapping) beam). 

Given $|k_{t}-k_{c}|\ll k_{c}$, one can regard $(k_{t}-k_{c})\hat{x}$
as a position-dependent phase shift between the standing waves of
the two intracavity fields via $\sin^{2}(k_{t}\hat{x})=\sin^{2}\left(k_{c}\hat{x}+(k_{t}-k_{c})\hat{x}\right)=\sin^{2}\left(k_{c}\hat{x}+\varphi\right)$,
where 
\[
\varphi=(k_{t}-k_{c})x_{0}^{\prime}=\frac{2\pi\text{{FSR}}}{c}x_{0}^{\prime}=\frac{\pi}{L}x_{0}^{\prime}
\]

We include the dependence on $k_{t}$ in $\varphi$ and use $k$ instead
of $k_{c}$ from this point on. Further, we rewrite the position operator
$\hat{x}$ as the sum of three terms: $\hat{x}=x_{0}^{\prime}+\bar{x}+\hat{x}_{m}$,
where $x_{0}^{\prime}$ is the position of the intensity maximum of
the control field with respect to the cavity mirror ($\bar{x}$: the
nanoparticle's mean displacement from $x_{0}^{\prime}$, $\hat{x}_{m}$:
the nanoparticle's displaced position operator with $\langle\hat{x}_{m}\rangle=0$).
Note that in the main text we always use the distance from the cavity
center $x_{0}$, where $x_{0}^{\prime}=x_{0}+L/2.$ We also introduce
the dimensionless position operator $\delta\hat{x}$ with $\hat{x}_{m}=X_{\text{\text{{gs}}}}\cdot\delta\hat{x}$,
where $\delta\hat{x}=\frac{1}{\sqrt{2}}(\hat{b}+\hat{b}^{\dagger})$
($X_{\text{{gs}}}$: Ground state extension of the mechanical oscillator,
$b^{(\dagger)}$: CM-motion annihilation (creation) operator).

We approximate the trigonometric functions in equation \ref{eq: Ham1}
to a second-order in $\hat{x}_{m}$ and perform a displacement operation
of the light operators: $\hat{a}_{j}\rightarrow\alpha_{j}+\hat{a}_{j}$
about their steady-state mean values $\alpha_{t}$ and $\alpha_{c}$.
The Hamiltonian after these modifications is: 
\begin{eqnarray}
\frac{H}{\hbar} & = & \Delta_{t}|\alpha_{t}|^{2}+\Delta_{c}|\alpha_{c}|^{2}+\Delta_{t}\alpha_{t}(\hat{a}_{t}+\hat{a}_{t}^{\dagger})+\Delta_{c}\alpha_{c}(\hat{a}_{c}+\hat{a}_{c}^{\dagger})+\Delta_{t}\hat{a}_{t}^{\dagger}\hat{a}_{t}+\Delta_{c}\hat{a}_{c}^{\dagger}\hat{a}_{c}\nonumber \\
 & + & \frac{\hat{p}_{m}^{2}}{2m\hbar}-U_{0}|\alpha_{t}|^{2}\sin^{2}(k(x_{0}^{\prime}+\bar{x})+\varphi)-U_{0}|\alpha_{c}|^{2}\sin^{2}(k(x_{0}^{\prime}+\bar{x}))\nonumber \\
 & - & 2U_{0}k^{2}|\alpha_{t}|^{2}\cos(2k(x_{0}^{\prime}+\bar{x})+2\varphi)\frac{\hat{x}_{m}^{2}}{2}-2U_{0}k^{2}|\alpha_{c}|^{2}\cos(2k(x_{0}^{\prime}+\bar{x}))\frac{\hat{x}_{m}^{2}}{2}\label{wm} \nonumber\\
 & - & U_{0}k\alpha_{t}\sin(2k(x_{0}^{\prime}+\bar{x})+2\varphi)(\hat{a}_{t}+\hat{a}_{t}^{\dagger})\hat{x}_{m}-U_{0}k\alpha_{c}\sin(2k(x_{0}^{\prime}+\bar{x}))(\hat{a}_{c}+\hat{a}_{c}^{\dagger})\hat{x}_{m}\label{g1}\nonumber\\
 & - & U_{0}|\alpha_{t}|^{2}k\sin(2k(x_{0}^{\prime}+\bar{x})+2\varphi)\hat{x}_{m}-U_{0}|\alpha_{c}|^{2}k\sin(2k(x_{0}^{\prime}+\bar{x}))\hat{x}_{m}\nonumber \\
 & - & U_{0}\alpha_{t}\sin^{2}(k(x_{0}^{\prime}+\bar{x})+\varphi)(\hat{a}_{t}+\hat{a}_{t}^{\dagger})-U_{0}\alpha_{c}\sin^{2}(k(x_{0}^{\prime}+\bar{x}))(\hat{a}_{c}+\hat{a}_{c}^{\dagger})\nonumber \nonumber\\
 & + & iE_{t}(\hat{a}_{t}^{\dagger}-\hat{a}_{t})+iE_{c}(\hat{a}_{c}^{\dagger}-\hat{a}_{c}).
\end{eqnarray}

Line \ref{wm} takes the form of a harmonic potential $\frac{m\Omega_{0}^{2}\hat{x}_{m}^{2}}{2\hbar}$
with mechanical frequency $\Omega_{0}$: 
\begin{equation}
\Omega_{0}^{2}=-\frac{2\hbar U_{0}k^{2}}{m}\left(|\alpha_{t}|^{2}\cos(2k(x_{0}^{\prime}+\bar{x})+2\varphi)+|\alpha_{c}|^{2}\cos(2k(x_{0}^{\prime}+\bar{x}))\right).\label{wmr}
\end{equation}

Line \ref{g1} determines the linear dispersive coupling of the nanosphere
CM motion to the trapping and cooling beam. Note that the trapping
beam also shows linear coupling when the cooling beam is strong enough
to significantly contribute to the optical trap: 
\begin{eqnarray}
g_{0,t} & = & \zeta_{t}\cdot X_{\text{{gs}}}=U_{0}kX_{\text{{gs}}}\sin2(k(x_{0}^{\prime}+\bar{x})+\varphi)\nonumber \\
g_{0,c} & = & \zeta_{c}\cdot X_{\text{{gs}}}=U_{0}kX_{\text{{gs}}}\sin2k(x_{0}^{\prime}+\bar{x}).\label{coupling}
\end{eqnarray}

Note that we use $g_{0}=g_{0,c}$ in the main text. To study the dynamics
of the system, we solve the Langevin equations for both light fields:
\begin{eqnarray}
\dot{\hat{a}}_{t} & = & -(\frac{\kappa}{2}+i(\Delta_{t}-U_{0}\sin^{2}(k(x_{0}^{\prime}+\bar{x})+\varphi))(\hat{a}_{t}+\alpha_{t})+E_{t}-\zeta_{t}\alpha_{t}\hat{x}_{m}\nonumber \\
\dot{\hat{a}}_{c} & = & -(\frac{\kappa}{2}+i(\Delta_{c}-U_{0}\sin^{2}k(x_{0}^{\prime}+\bar{x}))(\hat{a}_{c}+\alpha_{c})+E_{c}-\zeta_{c}\alpha_{c}\hat{x}_{m}\label{as}
\end{eqnarray}

The additional loss terms account for the cavity amplitude decay rate
$\kappa$. The value of $\kappa$ is assumed to be equal for both
light fields due to the small difference in their wavelengths. For
the steady-state solutions of $\hat{a}_{t}$ and $\hat{a}_{c}$ we
find: 
\begin{eqnarray}
\alpha_{t} & = & \frac{E_{t}}{\frac{\kappa}{2}+i(\Delta_{t}-U_{0}\sin^{2}k(x_{0}^{\prime}+\bar{x})+\varphi)}\nonumber \\
\alpha_{c} & = & \frac{E_{c}}{\frac{\kappa}{2}+i\left(\Delta_{c}-U_{0}\sin^{2}(k(x_{0}^{\prime}+\bar{x}))\right)}.\label{alphas}
\end{eqnarray}

In our experiment, the Pound-Drever-Hall feedback loop keeps the trapping-laser
frequency resonant to the corresponding cavity resonance frequency
when the particle is in its steady state position. In other words,
the detuning $\Delta_{t}$ compensates the frequency shift caused
by the particle such that $\Delta_{t}-U_{0}\sin^{2}(k(x_{0}^{\prime}+\bar{x})+\varphi)=0$. 

On the other hand, the frequency of the control beam is varied throughout
the experiment. We are interested in the detuning $\Delta$ of the
control beam with respect to the cavity resonance when the nanoparticle
is located at its steady state position: $\Delta=\Delta_{c}-U_{0}\sin^{2}k(x_{0}^{\prime}+\bar{x})$.

The trapping beam power is not changed throughout the experiment.
In contrast, the control beam power is always set to achieve the desired
ratio between the power of the two intracavity fields $\mu$: 
\begin{equation}
\mu=\frac{|\alpha_{c}|^{2}}{|\alpha_{t}|^{2}}=\frac{|E_{c}|^{2}}{|E_{t}|^{2}}\frac{\left(\frac{\kappa}{2}\right)^{2}}{\left(\frac{\kappa}{2}\right)^{2}+\Delta^{2}}.
\end{equation}

Heisenberg's equation of motion for the particle becomes: 
\begin{equation}
\ddot{\hat{x}}_{m}+\Omega_{0}^{2}(\mu)\hat{x}_{m}=\frac{\hbar kU_{0}}{m}\left[|\alpha_{t}|^{2}\sin2(k(x_{0}^{\prime}+\bar{x})+\varphi)+|\alpha_{c}|^{2}\sin2k(x_{0}^{\prime}+\bar{x})\right]-\gamma_{m}\dot{\hat{x}}_{m}\label{xH}
\end{equation}
where we included an additional damping term $\gamma_{m}\dot{\hat{x}}_{m}$,
which is due to the collisions of the nanoparticle with the surrounding
gas.

From Equation \ref{xH} we find a steady state condition
on $x_{0}^{\prime}+\bar{x}$, that enables us to determine the mechanical
frequency $\Omega_{0}$ and the displacement $\bar{x}$ as a function
of $\mu$: 
\begin{eqnarray}
\Omega_{0}^{2}(\mu) & = & \Omega_{0}^{2}(0)\sqrt{1+\mu^{2}+2\mu\cos2\varphi}\label{frequency_asaf_mu}\\
\tan2k\bar{x} & = & -\frac{\sin2\varphi}{\mu+\cos2\varphi}.\nonumber 
\end{eqnarray}

Thereby, the mechanical frequency in absence of the cooling beam is
(equation \ref{wmr}): 
\[
\Omega_{0}^{2}(0)=\frac{2\hbar U_{0}k^{2}}{m}|\alpha_{t}|^{2}.
\]

Note that the case of a control beam that significantly contributes
to the optical trap that has been presented here has also already
been published in \cite{Pender2012,Monteiro2013}.

\subsection*{Cavity mode shape}

Up to this point, we have neglected the mode shape of the TEM00 cavity
mode (Fig.~\ref{Figure 2}a, main text). The waist of the mode, however, depends
on the position $x_{0}$ in the cavity. The maximum intensity of the
standing wave along the TEM00 mode in the cavity is, accordingly,
position dependent \cite{saleh2007}: 
\[
I(x_{0})=I_{0}\frac{1}{1+\frac{x_{0}^{2}}{x_{R}^{2}}},
\]
note, that we have used here $x_{0}$ as the distance from the center
of the cavity. It is related to the distance from the mirror $x_{0}^{\prime}$
by $x_{0}=x_{0}^{\prime}-\frac{L}{2}$ ($x_{R}$: Rayleigh length
of the mode). Therefore, $U_{0}$ is an explicit function of the trap
position $x_{0}$: 
\[
U_{0}(x_{0})=\frac{\omega_{cav}\xi}{2\varepsilon_{0}V_{c}}\left(1+\frac{x_{0}^{2}}{x_{R}^{2}}\right)^{-1},
\]
($\omega_{cav}$: laser frequency, $\xi$: particle polarizability,
$\varepsilon_{0}$: vacuum permittivity, $V_{c}$: cavity mode volume).
The polarizability of a particle is (see e.g. \cite{Chang2010}):
\[
\xi=4\pi r^{3}\varepsilon_{0}\operatorname{Re}\left(\frac{\varepsilon-1}{\varepsilon+2}\right)
\]
($\varepsilon$: nanoparticle's dielectric constant; $r$: particle
radius). In the main text we use these equations to determine the
estimated particle size from $U_{0}(x_{0})$, which is determined
from the control beam power dependent coupling $g_{0}$(see main text,
figure \ref{Fig3}d) and the independently determied position of a particle
in the cavity $x_{0}$(see appendix \ref{app:position}).

\subsection*{Langevin Equations, effective frequency and damping}

The Langevin equations for the mechanical quantum harmonic oscillator
coupled to a thermal bath are: 
\begin{eqnarray}
\dot{\hat{x}}_{m} & = & \frac{\hat{p}_{m}}{m}\nonumber \\
\dot{\hat{p}}_{m} & = & -m\Omega_{0}^{2}\hat{x}_{m}-\gamma_{m}\hat{p}_{m}+\sum_{j=t,c}\hbar\zeta_{j}\alpha_{j}(\hat{a}_{j}^{\dagger}+\hat{a}_{j})+\eta(t),\label{mech}
\end{eqnarray}
where $\eta(t)$ is a thermal noise term, with the following correlation
property \cite{Genes2008a}: 
\[
\langle\eta(t)\eta(t')\rangle=\frac{\gamma_{m}}{\Omega_{0}}\int\frac{d\omega}{2\pi}e^{-i\omega(t-t')}\omega\coth\left(\frac{\hbar\omega}{2k_{B}T}\right).
\]
We assume that we are in a temperature range where $k_{B}T/\hbar\gg\Omega_{0}$:
\[
\langle\eta(t)\eta(t')\rangle=\gamma_{m}\frac{2k_{B}T}{\hbar\Omega_{0}}\delta(t-t').
\]

For the light beams, we can use the equations of motion as provided
in equation \ref{as} after the displacement of the light operators:
\begin{eqnarray}
\dot{\hat{a}}_{c} & = & -(\frac{\kappa}{2}+i\Delta)\hat{a}_{c}+i\zeta_{c}\alpha_{c}\hat{x}_{m}+\sqrt{\kappa}(\hat{c}_{c}^{in}+\hat{d}_{c}^{in}),\label{fouriercon}\\
\dot{\hat{a}}_{t} & = & -\frac{\kappa}{2}\hat{a}_{t}+i\zeta_{t}\alpha_{t}\hat{x}_{m}+\sqrt{\kappa}(\hat{c}_{t}^{in}+\hat{d}_{t}^{in})\nonumber 
\end{eqnarray}

By Fouriertransformation, we obtain a linear system of equations from
which we retrieve the final expression for the position spectrum $S_{xx}(\omega)$
of levitating nanoparticles CM motion: 
\[
S_{xx}=\left|\chi_{m}^{\text{eff}}\right|^{2}[S_{\text{th}}+S_{\text{rp}}],
\]
where $S_{\text{th}}$ is the thermal noise contribution and $S_{\text{rp}}$
is the radiation-pressure contribution. In the regime our experiment
is currently operating ($T=293$ K; air pressure approx. $1$-$5$
mbar), we expect that the thermal-noise contribution prevales: 
\[
S_{\text{th}}=X_{gs}^{2}\gamma_{m}\frac{2k_{B}T}{\hbar\Omega_{0}}
\]

The effective susceptibility of the mechanical oscillator is
\begin{eqnarray}
\chi_{m}^{\text{eff}}=\frac{\gamma_{m}}{(\Omega_{\text{eff}}(\omega){}^{2}-\omega^{2})^{2}-i\gamma_{\text{eff}}(\omega)^{2}\omega^{2}},\label{eq:Susz}
\end{eqnarray}

where, following \cite{Genes2008a}, we used the expressions:

\begin{eqnarray*}
\gamma_{\text{eff}}(\omega) & = & \gamma_{m}-\frac{4g_{0}^{2}|\alpha_{c}|^{2}\Omega_{0}(\mu)\Delta\frac{\kappa}{2}}{\left(\left(\frac{\kappa}{2}\right)^{2}+(\omega+\Delta)^{2}\right)\left(\left(\frac{\kappa}{2}\right)^{2}+(\omega-\Delta)^{2}\right)}\\
\Omega_{\text{eff}}(\omega) & = & \left[\Omega_{0}^{2}(\mu)+\frac{2g_{0}^{2}|\alpha_{c}|^{2}\Omega_{0}(\mu)\Delta\left[\left(\frac{\kappa}{2}\right)^{2}-\omega^{2}+\Delta^{2}\right]}{\left(\left(\frac{\kappa}{2}\right)^{2}+(\omega+\Delta)^{2}\right)\left(\left(\frac{\kappa}{2}\right)^{2}+(\omega-\Delta)^{2}\right)}\right]^{1/2}.
\end{eqnarray*}

\section{Position readout by homodyne detection of the control beam}
\label{app:readout}
The expressions for the mechanical oscillator's dynamics, as well
as its relationship to the control beam in the cavity, have been derived
in the previous section (equations \ref{mech} and \ref{fouriercon}).
In the following two sections, we will discuss how the mechanical
oscillator position NPS is determined from the NPS obtained by homodyning
of the control-beam phase signal in transmission of the cavity (see
Methods M2 for implementation of homodyne detection). 

We first derive the control light field in cavity transmission via
the cavity input-output relation \cite{gardiner2004}:

\begin{equation}
\hat{d}_{c}^{out}(t)=\sqrt{\kappa}\hat{a}_{c}(t)-\hat{d}_{c}^{in}(t),\label{inout}
\end{equation}

where $\hat{d}_{c}^{in}$ describes the quantum noise at the cavity
back mirror (i.e., the side from which the cavity is not driven).
Even though our detection scheme occurs in two steps as described
in the methods section, it is completely equivalent to a standard
homodyne detection. The output signal is accordingly described by
\cite{bachor2004guide}: 
\begin{equation}
s_{\text{opt}}(t)=\frac{1}{2}\left(|\hat{d}_{c}^{out}+\hat{a}_{LO}|^{2}-|\hat{d}_{c}^{out}-\hat{a}_{LO}|^{2}\right)=\hat{d}_{c}^{out}\hat{a}_{LO}^{*}+\hat{d}_{c}^{out\dagger}\hat{a}_{LO},\label{sig}
\end{equation}
where we describe the local oscillator by $\hat{a}_{LO}(t)=\alpha_{LO}\cdot e^{-i(\omega_{c}t+\theta)}$,
where $\theta$ determines the detected quadrature of the control
beam and $\alpha_{LO}$ is assumed to be real. In our experiment,
the readout phase is locked to measure the phase quadrature: $\theta=\frac{\pi}{2}$. 

From equations \ref{inout} and \ref{sig} we obtain: 
\begin{equation}
\langle|\tilde{s}_{\text{opt}}(\omega)|^{2}\rangle=\kappa\zeta_{c}^{2}\alpha_{c}^{2}\left|\chi_{c}(\omega)+\chi_{c}^{*}(-\omega)\right|^{2}S_{xx}(\omega)\delta(\omega),\label{spec}
\end{equation}
where we used the cavity susceptibility $\chi_{c}(\omega)=\frac{1}{\frac{\kappa}{2}-i(\omega-\Delta)}$.
Up to a proportionality factor, Equation \ref{spec} resembles the
result of the detection described in \cite{Paternostro2006}, and
allows us to derive the mechanical NPS $S_{xx}(\omega)$ from the
detected signal.

\section{Data Evaluation and Temperature Calibration }
\label{app:evaluation}
To extract the mechanical NPS, we first measure the spectrum of the
homodyne phase readout with and without particle for all values of
$\mu$ and $\Delta$. We obtain $\langle|\tilde{s}_{\text{opt}}(\omega)|^{2}\rangle$
by substracting the background NPS (without particle) from the NPS
with particle. To reconstruct the mechanical NPS $S_{xx}$, we need
to account for the filtering by the Fabry-Perot cavity. We therefore
divide $\langle|\tilde{s}_{\text{opt}}(\omega)|^{2}\rangle$ by $|\chi_{c}(\omega)+\chi_{c}^{*}(-\omega)|^{2}$
following equation \ref{spec}. The exact shape of $S_{xx}$ is given
by equation \ref{eq:Susz}. To determine the effective frequency,
damping and temperature we assume that we can describe the CM-motion
of the particle as an harmonic oscillator, which is fulfilled as we
are not operating in the strong coupling regime:

\begin{figure}[ht]
\includegraphics[width=0.6\linewidth]{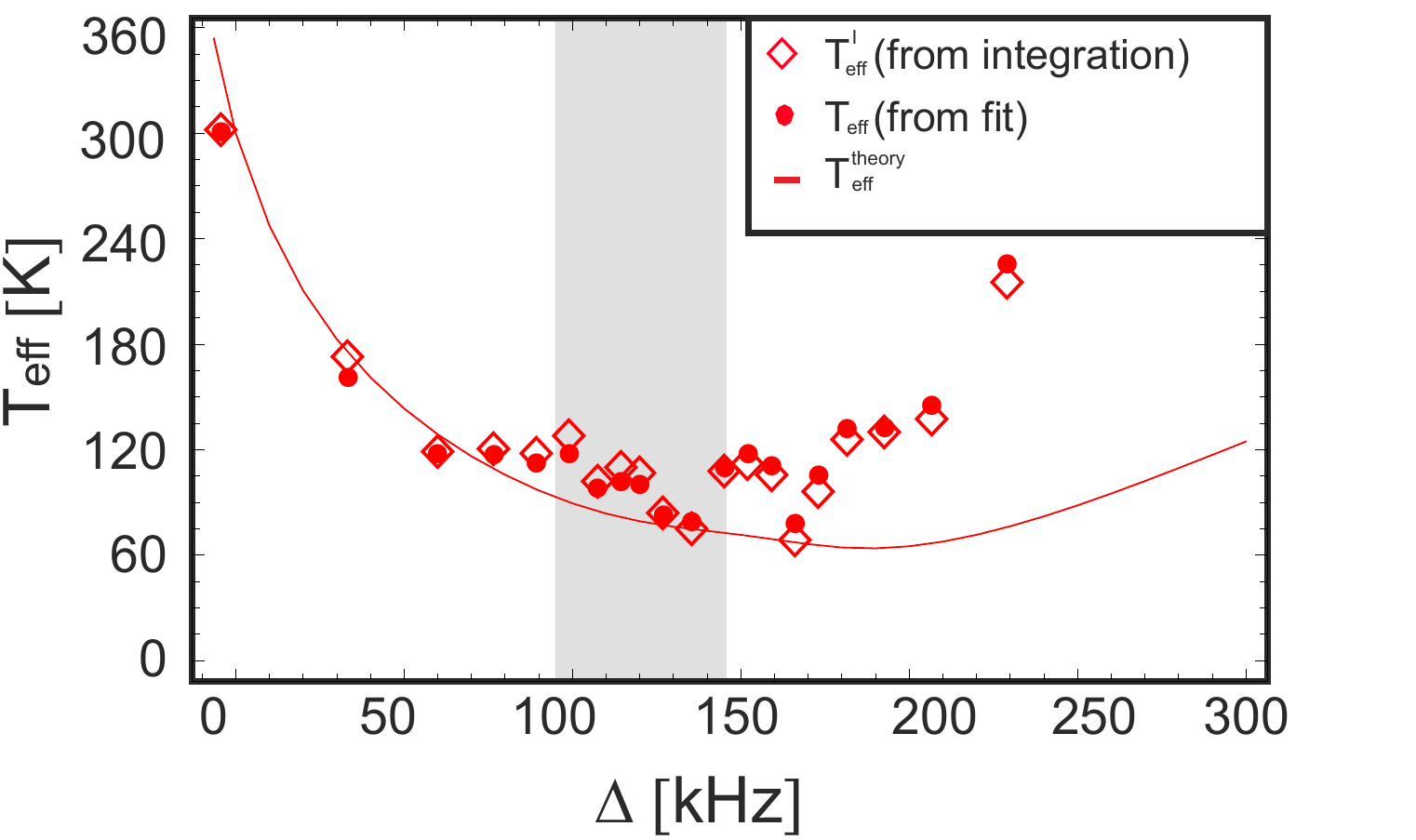}\caption{\textbf{CM-motion temperature as a function of detuning.} 
\label{coolingplot}
The temperature
of the CM-motion along the cavity axis as a function of detuning.
The values are inferred via the equipartition theorem from the direct
integration of the NPS ($T_{\textrm{eff}}^{I},$solid circles) and
from the fitted spectra ($T_{\textrm{eff}}^{*}$, empty diamonds).
The solid line shows the theoretical expectation $T_{\textrm{eff}}^{\text{{theory}}}(\Delta)$
infered from the detuning dependent frequency fit (optical spring;
main text, Fig.~\ref{Fig3}).}
\end{figure}

\begin{equation}
f(\omega)=a\cdot T_{\textrm{eff}}\cdot\frac{\gamma_{\textrm{m}}}{(\omega^{2}-\Omega_{\text{eff}}^{2})^{2}+\omega^{2}\gamma_{\text{eff}}^{2}}.\label{fitmodel}
\end{equation}

By fitting this model to $S_{xx}$, we obtain $\gamma_{\textrm{eff}}$,
$\Omega_{\text{eff}}$ and $T_{\textrm{eff}}^{*}$. The calibration
constant $a$ is determined such that $T_{\textrm{eff}}=293$K in
a particular measurement that was performed close to zero detuning
($\Delta=1$ kHz for $\mu=0.4$, blue NPS in Fig.~\ref{Fig3}, main text).
This results in the values for the optical spring $\Omega_{\text{eff}}$
and damping $\gamma_{\textrm{eff}}$ in Fig.~\ref{Fig3}, main text.

We can determine the optomechanical coupling $g_{0}$ from the the
detuning dependence of $\Omega_{\text{eff}}$ for a given value of
$\mu$. However, we do not have an explicit analytical expression
for this dependence. Instead, we apply the following strategy: 

Using equation \ref{eq:Susz}, we can calculate the optomechanical
NPS $S_{xx}^{\text{{theory}}}$ of our system for a given set of parameters
($\kappa$, $g_{0}$, $\Omega_{0}$ and $\delta\Delta$). Here, $\delta\Delta$
is a systematic deviation from the detuning we set in the measurement:
each value of $\Delta$ can be set precisely up to the uncertainty
in the actual cavity resonance frequency. This frequency difference
is accounted for with a joint offset $\delta\Delta$ in the values
of $\Delta$ that is used as a fit parameter. We treat $S_{xx}^{\text{{theory}}}$
in the same manner as the data and extract $\gamma_{\textrm{eff}}^{\text{{theory}}}$,
$\Omega_{\text{eff}}^{\text{{theory}}}$ and $T_{\textrm{eff}}^{\text{{theory}}}$
by fitting $f(\omega)$ for each value of $\Delta$. We use $\Omega_{\text{eff}}^{\text{{theory}}}(\Delta)$
as a model that we fit to $\Omega{}_{\text{eff}}$ optimizing the
parameters $g_{0}$, $\Omega_{0}$ and $\delta\omega$ in a least-square
fit. The FWHM cavity line width $\kappa$ is determined independently.
The best fit parameters are used to obtain the theoretical dependences
of $\gamma_{\textrm{eff}}^{\text{{theory}}}$ and $\Omega_{\text{eff}}^{\text{{theory}}}$
on the detuning shown in Fig.~\ref{Fig3} in the main text and $T_{\text{eff}}^{\text{theory}}$
shown in Fig.~\ref{coolingplot}.

The corresponding values of the predicted effective temperatures $T_{\textrm{eff}}^{\text{{theory}}}$
are shown in Fig.~\ref{coolingplot} along with the experimental data
for $\mu=0.4$. The latter is obtained in two ways: firstly as a free
parameter $T_{\textrm{eff}}^{*}$ in the fitted model $f(\omega)$
and secondly by direct integration over the measured NPS via $T_{\textrm{eff}}^{I}=a^{I}\Omega_{\textrm{eff}}^{2}\int S_{xx}d\omega$.
The calibration factor $a^{I}$ is derived in the same way as $a$.
The values $T_{\textrm{eff}}^{*}$, obtained via fitting, agree well
with those obtained by direct integration of the NPS. For small detunings
$\Delta$, the data follows the theoretical curve, while for larger
detunings, heating unaccounted for in the theoretical model seems
to occur. We are still investigating this effect, which may be due
to laser noise. To obtain a good estimate of the minimal temperature
achieved experimentally, we average the temperature obtained for a
range of detunings $\Delta/2\pi\in[100,150]$ kHz. The range is chosen
such that the onset of temperature increase is not yet strong and
the predicted range of $T_{\textrm{eff}}$ is small compared to the
distribution of measured temperatures. The experimental data in Fig.~\ref{Fig3}e in the main text is obtained by applying this evaluation for
the different values of $\mu$ for $T_{\textrm{eff}}^{I}$ obtained
by direct integration. The theory curve in Fig.~\ref{Fig3}e in the main text
is obtained by averaging the theoretical prediction for $T_{\textrm{eff}}^{\text{theory}}$
over the same range of detunings $\Delta/2\pi\in[100,150]$ kHz.

\begin{figure}
\includegraphics[width=0.7\linewidth]{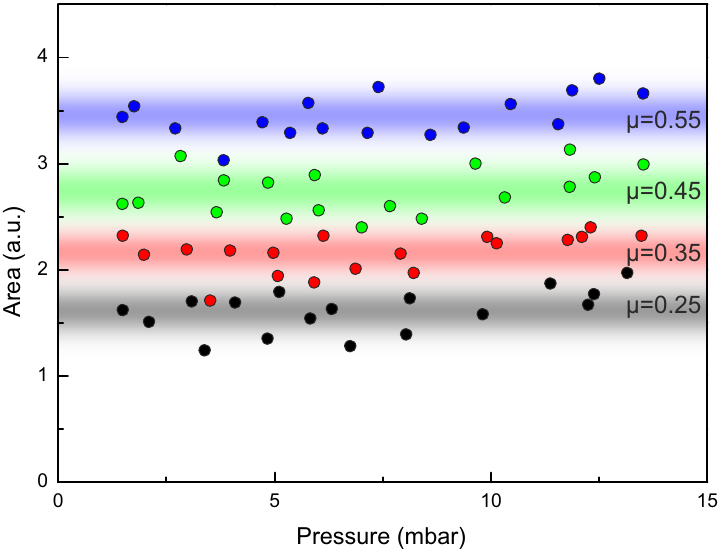}\caption{Measurement of the NPS-area as a function of the ambient pressure.
\label{fig_pressure_2} 
The data were taken for a resonant control beam. The area is proportional
to the temperature of the CM motion of the nanoparticle for a given
value of $\mu$. It is independent of pressure indicating that the
CM motion of the trapped nanoparticles is thermalized with the surrounding
gas over the full measurement range. We conclude that the effective
temperature of the CM mode is room temperature ($293$ K). The scatter
of the data corresponds to an standard deviation of $5$\% for a temperature
that is infered from the area of the NPS.}
\end{figure}

\section{Kinetic gas theory - Pressure-dependent damping }
\label{app:gas}
The pressure dependence of the damping for a trapped nanosphere is
given by \cite{Beresnev1990,Li2011b,Gieseler2012} : 
\begin{equation}
\gamma_{0}=\frac{6\pi\eta r}{m}\frac{0.619}{0.619+Kn}(1+c_{\text{k}})\label{eq:pressuredamp}
\end{equation}
where $\eta$ is the viscosity coefficient for air, $r$ and $m$
are the radius and mass of the nanosphere, $Kn=\lambda_{\text{fp}}/r$
is the Knudsen number and $\lambda_{\text{fp}}$ is the mean free
path for air particles. $c_{\text{k}}=0.31Kn/(0.785+1.152Kn+Kn^{2})$
is a small correction factor necessary at higher pressures \cite{Beresnev1990}.
Figure \ref{Figure 2}b in the main text shows a pressure dependent damping measurement,
where the control beam is used just for readout (i.e. $\mu$=0.1,
resonant). 

Figure \ref{fig_pressure_2} shows the temperature associated with
the CM motion of the particle as a function of pressure. At high pressures,
the nanoparticle experiences more collisions with the gas resulting
in a stronger damping of its CM motion. If the nanoparticle CM motion
was not thermalized at low pressures due to a heating process, better
thermalization and therefore lower temperatures would be expected
at higher pressures due to the increased damping rate. As Fig.~\ref{fig_pressure_2}
shows a constant CM motion temperature for the different pressures
and values of $\mu$, we conclude it is thermalized with the environment
in all these measurements, which implies a temperature of 293K as
long as no optical damping is introduced.

\begin{figure}[ht]
\includegraphics[width=0.4\linewidth]{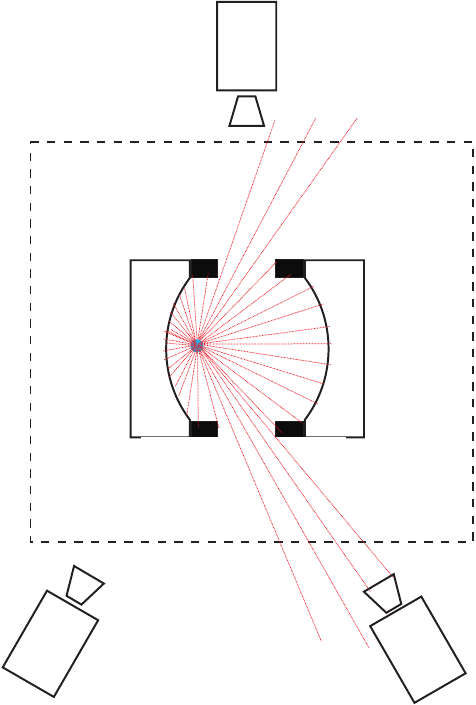}\caption{\textbf{Schematic of the configuration of the CCD imaging setup}.
\label{cameras}
We use a combination of three CCD cameras to observe and locate the
particle. This configuration allows access to a larger range of positions
along the cavity axis compared to a single camera. Based on the pictures
of the CCD cameras, we determine the position $x_{0}$ of a particle
in the cavity.}
\end{figure}

\section{Position detection}
\label{app:position}
Three cameras with achromatic lenses monitor the cavity and image
the light scattered off trapped nanoparticles. As can be seen in Figure \ref{Fig1}a of the main text, the black retaining rings and the concave shape
of the mirrors prevent optical access over the whole cavity length
from a single point of view. We use a configuration of three CCD cameras,
as shown in figure~\ref{cameras}, to extend the field of view. By
combining the images from the 3 cameras, we can reconstruct a larger
field of view. To determine the position of the particle from the
image, we need to calibrate the coordinates. To this end, the mechanical
frequency at several positions is measured along with the position
of a particle on the CCD image. This measurement is repeated for several
particles. The frequency dependence on position allows calibrating
the camera. 

The mode shape in the optical cavity is well-known from the curvature
of the mirrors and the cavity length, which is determined with high
precision from the FSR. The expected longitudinal frequency dependence
of a nanoparticle trapped in the standing wave is $\Omega_{0}=\Omega_{c}\frac{1}{\sqrt{1+((x_{0}-x_{c})/x_{R})^{2}}}$
($x_{c}$: cavity center position, $\Omega_{c}$: frequency at position
$x_{c}$; $x_{R}$: Rayleigh length of the Gaussian mode). The measured
mechanical frequencies for several different trap positions of the
same nanoparticle and the corresponding coordinates $\zeta$ (in pixels)
on the camera images are fitted to the function $\Omega_{0}=\Omega_{c}\frac{1}{\sqrt{1+((\zeta-\zeta_{c})\xi/x_{R})^{2}}}$
with fit parameters $\zeta_{c}$ (coordinate of the center of the
cavity in pixels), $\xi$ (conversion factor between pixels and millimeters)
and $\Omega_{c}$ (mechanical frequency in the center of the cavity).
A corresponding measurement for one nanoparticle with calibrated length
scale is shown in the main text, figure \ref{Figure 2}b. Based on this calibration
we determine the position of the nanoparticle used in the measurements
summarized in Fig.~\ref{Fig3}. It is located at a distance $x_{0}^{\prime}=3.92\pm0.14$
mm from the cavity mirror, i.e. at a distance $x_{0}=1.56\pm0.14$
mm from the center of the cavity.

\bibliography{LevitationbibArxiv}

\begin{thebibliography}{59}
\expandafter\ifx\csname natexlab\endcsname\relax\def\natexlab#1{#1}\fi
\expandafter\ifx\csname bibnamefont\endcsname\relax
  \def\bibnamefont#1{#1}\fi
\expandafter\ifx\csname bibfnamefont\endcsname\relax
  \def\bibfnamefont#1{#1}\fi
\expandafter\ifx\csname citenamefont\endcsname\relax
  \def\citenamefont#1{#1}\fi
\expandafter\ifx\csname url\endcsname\relax
  \def\url#1{\texttt{#1}}\fi
\expandafter\ifx\csname urlprefix\endcsname\relax\def\urlprefix{URL }\fi
\providecommand{\bibinfo}[2]{#2}
\providecommand{\eprint}[2][]{\url{#2}}

\bibitem[{\citenamefont{Blatt and Roos}(2012)}]{Blatt2012}
\bibinfo{author}{\bibfnamefont{R.}~\bibnamefont{Blatt}} \bibnamefont{and}
  \bibinfo{author}{\bibfnamefont{C.~F.} \bibnamefont{Roos}},
  \bibinfo{journal}{Nature Physics} \textbf{\bibinfo{volume}{8}},
  \bibinfo{pages}{277} (\bibinfo{year}{2012}).

\bibitem[{\citenamefont{Bloch et~al.}(2012)\citenamefont{Bloch, Dalibard, and
  Nascimb\`{e}ne}}]{Bloch2012}
\bibinfo{author}{\bibfnamefont{I.}~\bibnamefont{Bloch}},
  \bibinfo{author}{\bibfnamefont{J.}~\bibnamefont{Dalibard}}, \bibnamefont{and}
  \bibinfo{author}{\bibfnamefont{S.}~\bibnamefont{Nascimb\`{e}ne}},
  \bibinfo{journal}{Nature Physics} \textbf{\bibinfo{volume}{8}},
  \bibinfo{pages}{267} (\bibinfo{year}{2012}).

\bibitem[{\citenamefont{Haffner et~al.}(2008)\citenamefont{Haffner, C., and
  Blatt}}]{HAFFNER2008}
\bibinfo{author}{\bibfnamefont{H.}~\bibnamefont{Haffner}},
  \bibinfo{author}{\bibfnamefont{R.}~\bibnamefont{C.}}, \bibnamefont{and}
  \bibinfo{author}{\bibfnamefont{R.}~\bibnamefont{Blatt}},
  \bibinfo{journal}{Physics Reports} \textbf{\bibinfo{volume}{469}},
  \bibinfo{pages}{155} (\bibinfo{year}{2008}).

\bibitem[{\citenamefont{Kielpinski et~al.}(2002)\citenamefont{Kielpinski,
  Monroe, and Wineland}}]{Kielpinski2002}
\bibinfo{author}{\bibfnamefont{D.}~\bibnamefont{Kielpinski}},
  \bibinfo{author}{\bibfnamefont{C.}~\bibnamefont{Monroe}}, \bibnamefont{and}
  \bibinfo{author}{\bibfnamefont{D.~J.} \bibnamefont{Wineland}},
  \bibinfo{journal}{Nature} \textbf{\bibinfo{volume}{417}},
  \bibinfo{pages}{709} (\bibinfo{year}{2002}).

\bibitem[{\citenamefont{Kimble}(2008)}]{Kimble2008}
\bibinfo{author}{\bibfnamefont{H.~J.} \bibnamefont{Kimble}},
  \bibinfo{journal}{Nature} \textbf{\bibinfo{volume}{453}},
  \bibinfo{pages}{1023} (\bibinfo{year}{2008}).

\bibitem[{\citenamefont{Ritter et~al.}(2012)\citenamefont{Ritter, N\"{o}lleke,
  Hahn, Reiserer, Neuzner, Uphoff, M\"{u}cke, Figueroa, Bochmann, and
  Rempe}}]{Ritter2012}
\bibinfo{author}{\bibfnamefont{S.}~\bibnamefont{Ritter}},
  \bibinfo{author}{\bibfnamefont{C.}~\bibnamefont{N\"{o}lleke}},
  \bibinfo{author}{\bibfnamefont{C.}~\bibnamefont{Hahn}},
  \bibinfo{author}{\bibfnamefont{A.}~\bibnamefont{Reiserer}},
  \bibinfo{author}{\bibfnamefont{A.}~\bibnamefont{Neuzner}},
  \bibinfo{author}{\bibfnamefont{M.}~\bibnamefont{Uphoff}},
  \bibinfo{author}{\bibfnamefont{M.}~\bibnamefont{M\"{u}cke}},
  \bibinfo{author}{\bibfnamefont{E.}~\bibnamefont{Figueroa}},
  \bibinfo{author}{\bibfnamefont{J.}~\bibnamefont{Bochmann}}, \bibnamefont{and}
  \bibinfo{author}{\bibfnamefont{G.}~\bibnamefont{Rempe}},
  \bibinfo{journal}{Nature} \textbf{\bibinfo{volume}{484}},
  \bibinfo{pages}{195} (\bibinfo{year}{2012}).

\bibitem[{\citenamefont{Stute et~al.}(2013)\citenamefont{Stute, Casabone,
  Brandstatter, Friebe, E., and Blatt}}]{Stute2013}
\bibinfo{author}{\bibfnamefont{A.}~\bibnamefont{Stute}},
  \bibinfo{author}{\bibfnamefont{B.}~\bibnamefont{Casabone}},
  \bibinfo{author}{\bibfnamefont{B.}~\bibnamefont{Brandstatter}},
  \bibinfo{author}{\bibfnamefont{K.}~\bibnamefont{Friebe}},
  \bibinfo{author}{\bibfnamefont{T.}~\bibnamefont{E.}, \bibfnamefont{Northup}},
  \bibnamefont{and} \bibinfo{author}{\bibfnamefont{R.}~\bibnamefont{Blatt}},
  \bibinfo{journal}{Nat Photon} \textbf{\bibinfo{volume}{7}},
  \bibinfo{pages}{219} (\bibinfo{year}{2013}).

\bibitem[{\citenamefont{Hofmann et~al.}(2012)\citenamefont{Hofmann, Krug,
  Ortegel, G\'{e}rard, Weber, Rosenfeld, and Weinfurter}}]{Hofmann2012}
\bibinfo{author}{\bibfnamefont{J.}~\bibnamefont{Hofmann}},
  \bibinfo{author}{\bibfnamefont{M.}~\bibnamefont{Krug}},
  \bibinfo{author}{\bibfnamefont{N.}~\bibnamefont{Ortegel}},
  \bibinfo{author}{\bibfnamefont{L.}~\bibnamefont{G\'{e}rard}},
  \bibinfo{author}{\bibfnamefont{M.}~\bibnamefont{Weber}},
  \bibinfo{author}{\bibfnamefont{W.}~\bibnamefont{Rosenfeld}},
  \bibnamefont{and}
  \bibinfo{author}{\bibfnamefont{H.}~\bibnamefont{Weinfurter}},
  \bibinfo{journal}{Science} \textbf{\bibinfo{volume}{337}},
  \bibinfo{pages}{72} (\bibinfo{year}{2012}).

\bibitem[{\citenamefont{Gerlich et~al.}(2011)\citenamefont{Gerlich,
  Eibenberger, Tomandl, Nimmrichter, Hornberger, Fagan, T\"{u}xen, Mayor, and
  Arndt}}]{Gerlich2011}
\bibinfo{author}{\bibfnamefont{S.}~\bibnamefont{Gerlich}},
  \bibinfo{author}{\bibfnamefont{S.}~\bibnamefont{Eibenberger}},
  \bibinfo{author}{\bibfnamefont{M.}~\bibnamefont{Tomandl}},
  \bibinfo{author}{\bibfnamefont{S.}~\bibnamefont{Nimmrichter}},
  \bibinfo{author}{\bibfnamefont{K.}~\bibnamefont{Hornberger}},
  \bibinfo{author}{\bibfnamefont{P.~J.} \bibnamefont{Fagan}},
  \bibinfo{author}{\bibfnamefont{J.}~\bibnamefont{T\"{u}xen}},
  \bibinfo{author}{\bibfnamefont{M.}~\bibnamefont{Mayor}}, \bibnamefont{and}
  \bibinfo{author}{\bibfnamefont{M.}~\bibnamefont{Arndt}},
  \bibinfo{journal}{Nature Communications} \textbf{\bibinfo{volume}{2}},
  \bibinfo{pages}{263} (\bibinfo{year}{2011}).

\bibitem[{\citenamefont{Hornberger et~al.}(2012)\citenamefont{Hornberger,
  Gerlich, Haslinger, Nimmrichter, and Arndt}}]{Hornberger2012}
\bibinfo{author}{\bibfnamefont{K.}~\bibnamefont{Hornberger}},
  \bibinfo{author}{\bibfnamefont{S.}~\bibnamefont{Gerlich}},
  \bibinfo{author}{\bibfnamefont{P.}~\bibnamefont{Haslinger}},
  \bibinfo{author}{\bibfnamefont{S.}~\bibnamefont{Nimmrichter}},
  \bibnamefont{and} \bibinfo{author}{\bibfnamefont{M.}~\bibnamefont{Arndt}},
  \bibinfo{journal}{Reviews of Modern Physics} \textbf{\bibinfo{volume}{84}},
  \bibinfo{pages}{157} (\bibinfo{year}{2012}).

\bibitem[{\citenamefont{Aspelmeyer et~al.}(2012)\citenamefont{Aspelmeyer,
  Meystre, and Schwab}}]{Aspelmeyer2012}
\bibinfo{author}{\bibfnamefont{M.}~\bibnamefont{Aspelmeyer}},
  \bibinfo{author}{\bibfnamefont{P.}~\bibnamefont{Meystre}}, \bibnamefont{and}
  \bibinfo{author}{\bibfnamefont{K.}~\bibnamefont{Schwab}},
  \bibinfo{journal}{Physics Today} \textbf{\bibinfo{volume}{65}},
  \bibinfo{pages}{29} (\bibinfo{year}{2012}).

\bibitem[{\citenamefont{Aspelmeyer et~al.}(2013)\citenamefont{Aspelmeyer,
  Kippenberg, and Marquardt}}]{Aspelmeyer2013}
\bibinfo{author}{\bibfnamefont{M.}~\bibnamefont{Aspelmeyer}},
  \bibinfo{author}{\bibfnamefont{T.}~\bibnamefont{Kippenberg}},
  \bibnamefont{and}
  \bibinfo{author}{\bibfnamefont{F.}~\bibnamefont{Marquardt}},
  \bibinfo{journal}{arXiv:1303.0733v1}  (\bibinfo{year}{2013}).

\bibitem[{\citenamefont{Chang et~al.}(2010)\citenamefont{Chang, Regal, Papp,
  Wilson, Ye, Painter, Kimble, and Zoller}}]{Chang2010}
\bibinfo{author}{\bibfnamefont{D.~E.} \bibnamefont{Chang}},
  \bibinfo{author}{\bibfnamefont{C.~A.} \bibnamefont{Regal}},
  \bibinfo{author}{\bibfnamefont{S.~B.} \bibnamefont{Papp}},
  \bibinfo{author}{\bibfnamefont{D.~J.} \bibnamefont{Wilson}},
  \bibinfo{author}{\bibfnamefont{J.}~\bibnamefont{Ye}},
  \bibinfo{author}{\bibfnamefont{O.}~\bibnamefont{Painter}},
  \bibinfo{author}{\bibfnamefont{H.~J.} \bibnamefont{Kimble}},
  \bibnamefont{and} \bibinfo{author}{\bibfnamefont{P.}~\bibnamefont{Zoller}},
  \bibinfo{journal}{Proceedings of the National Academy of Sciences of the
  United States of America} \textbf{\bibinfo{volume}{107}},
  \bibinfo{pages}{1005} (\bibinfo{year}{2010}).

\bibitem[{\citenamefont{Romero-Isart et~al.}(2010)\citenamefont{Romero-Isart,
  Juan, Quidant, and Cirac}}]{Romero-Isart2010}
\bibinfo{author}{\bibfnamefont{O.}~\bibnamefont{Romero-Isart}},
  \bibinfo{author}{\bibfnamefont{M.~L.} \bibnamefont{Juan}},
  \bibinfo{author}{\bibfnamefont{R.}~\bibnamefont{Quidant}}, \bibnamefont{and}
  \bibinfo{author}{\bibfnamefont{J.~I.} \bibnamefont{Cirac}},
  \bibinfo{journal}{New Journal of Physics} \textbf{\bibinfo{volume}{12}},
  \bibinfo{pages}{033015} (\bibinfo{year}{2010}).

\bibitem[{\citenamefont{Romero-Isart
  et~al.}(2011{\natexlab{a}})\citenamefont{Romero-Isart, Pflanzer, Blaser,
  Kaltenbaek, Kiesel, Aspelmeyer, and Cirac}}]{Romero-Isart2011a}
\bibinfo{author}{\bibfnamefont{O.}~\bibnamefont{Romero-Isart}},
  \bibinfo{author}{\bibfnamefont{A.~C.} \bibnamefont{Pflanzer}},
  \bibinfo{author}{\bibfnamefont{F.}~\bibnamefont{Blaser}},
  \bibinfo{author}{\bibfnamefont{R.}~\bibnamefont{Kaltenbaek}},
  \bibinfo{author}{\bibfnamefont{N.}~\bibnamefont{Kiesel}},
  \bibinfo{author}{\bibfnamefont{M.}~\bibnamefont{Aspelmeyer}},
  \bibnamefont{and} \bibinfo{author}{\bibfnamefont{J.~I.} \bibnamefont{Cirac}},
  \bibinfo{journal}{PRL} \textbf{\bibinfo{volume}{107}},
  \bibinfo{pages}{020405} (\bibinfo{year}{2011}{\natexlab{a}}).

\bibitem[{\citenamefont{Geraci et~al.}(2010)\citenamefont{Geraci, Papp, and
  Kitching}}]{Geraci2010}
\bibinfo{author}{\bibfnamefont{A.}~\bibnamefont{Geraci}},
  \bibinfo{author}{\bibfnamefont{S.}~\bibnamefont{Papp}}, \bibnamefont{and}
  \bibinfo{author}{\bibfnamefont{J.}~\bibnamefont{Kitching}},
  \bibinfo{journal}{Phys. Rev. Lett.} \textbf{\bibinfo{volume}{105}},
  \bibinfo{pages}{101101} (\bibinfo{year}{2010}).

\bibitem[{\citenamefont{Arvanitaki and Geraci}(2013)}]{Arvanitaki2013}
\bibinfo{author}{\bibfnamefont{A.}~\bibnamefont{Arvanitaki}} \bibnamefont{and}
  \bibinfo{author}{\bibfnamefont{A.~A.} \bibnamefont{Geraci}},
  \bibinfo{journal}{Phys. Rev. Lett.} \textbf{\bibinfo{volume}{110}},
  \bibinfo{pages}{071105} (\bibinfo{year}{2013}).

\bibitem[{\citenamefont{Horak et~al.}(1997)\citenamefont{Horak, Hechenblaikner,
  Gheri, Stecher, and Ritsch}}]{Horak1997a}
\bibinfo{author}{\bibfnamefont{P.}~\bibnamefont{Horak}},
  \bibinfo{author}{\bibfnamefont{G.}~\bibnamefont{Hechenblaikner}},
  \bibinfo{author}{\bibfnamefont{K.}~\bibnamefont{Gheri}},
  \bibinfo{author}{\bibfnamefont{H.}~\bibnamefont{Stecher}}, \bibnamefont{and}
  \bibinfo{author}{\bibfnamefont{H.}~\bibnamefont{Ritsch}},
  \bibinfo{journal}{Physical Review Letters} \textbf{\bibinfo{volume}{79}},
  \bibinfo{pages}{4974} (\bibinfo{year}{1997}).

\bibitem[{\citenamefont{Vuleti\'{c} and Chu}(2000)}]{Vuletic2000a}
\bibinfo{author}{\bibfnamefont{V.}~\bibnamefont{Vuleti\'{c}}} \bibnamefont{and}
  \bibinfo{author}{\bibfnamefont{S.}~\bibnamefont{Chu}},
  \bibinfo{journal}{Physical Review Letters} \textbf{\bibinfo{volume}{84}},
  \bibinfo{pages}{3787} (\bibinfo{year}{2000}).

\bibitem[{\citenamefont{Ashkin and Dziedzic}(1977)}]{Ashkin77}
\bibinfo{author}{\bibfnamefont{A.}~\bibnamefont{Ashkin}} \bibnamefont{and}
  \bibinfo{author}{\bibfnamefont{J.~M.} \bibnamefont{Dziedzic}},
  \bibinfo{journal}{Applied Physics Letters} \textbf{\bibinfo{volume}{30}},
  \bibinfo{pages}{202} (\bibinfo{year}{1977}).

\bibitem[{\citenamefont{Li et~al.}(2011)\citenamefont{Li, Kheifets, and
  Raizen}}]{Li2011b}
\bibinfo{author}{\bibfnamefont{T.}~\bibnamefont{Li}},
  \bibinfo{author}{\bibfnamefont{S.}~\bibnamefont{Kheifets}}, \bibnamefont{and}
  \bibinfo{author}{\bibfnamefont{M.~G.} \bibnamefont{Raizen}},
  \bibinfo{journal}{Nat Phys} \textbf{\bibinfo{volume}{7}},
  \bibinfo{pages}{527} (\bibinfo{year}{2011}).

\bibitem[{\citenamefont{Gieseler et~al.}(2012)\citenamefont{Gieseler, Deutsch,
  Quidant, and Novotny}}]{Gieseler2012}
\bibinfo{author}{\bibfnamefont{J.}~\bibnamefont{Gieseler}},
  \bibinfo{author}{\bibfnamefont{B.}~\bibnamefont{Deutsch}},
  \bibinfo{author}{\bibfnamefont{R.}~\bibnamefont{Quidant}}, \bibnamefont{and}
  \bibinfo{author}{\bibfnamefont{L.}~\bibnamefont{Novotny}},
  \bibinfo{journal}{Physical Review Letters} \textbf{\bibinfo{volume}{109}},
  \bibinfo{pages}{103603} (\bibinfo{year}{2012}).

\bibitem[{\citenamefont{Koch et~al.}(2010)\citenamefont{Koch, Sames, Kubanek,
  Apel, Balbach, Ourjoumtsev, Pinkse, and Rempe}}]{Koch2010}
\bibinfo{author}{\bibfnamefont{M.}~\bibnamefont{Koch}},
  \bibinfo{author}{\bibfnamefont{C.}~\bibnamefont{Sames}},
  \bibinfo{author}{\bibfnamefont{A.}~\bibnamefont{Kubanek}},
  \bibinfo{author}{\bibfnamefont{M.}~\bibnamefont{Apel}},
  \bibinfo{author}{\bibfnamefont{M.}~\bibnamefont{Balbach}},
  \bibinfo{author}{\bibfnamefont{A.}~\bibnamefont{Ourjoumtsev}},
  \bibinfo{author}{\bibfnamefont{P.}~\bibnamefont{Pinkse}}, \bibnamefont{and}
  \bibinfo{author}{\bibfnamefont{G.}~\bibnamefont{Rempe}},
  \bibinfo{journal}{Physical Review Letters} \textbf{\bibinfo{volume}{105}},
  \bibinfo{pages}{173003} (\bibinfo{year}{2010}).

\bibitem[{\citenamefont{Ye et~al.}(1999)\citenamefont{Ye, Vernooy, and
  Kimble}}]{Ye1999}
\bibinfo{author}{\bibfnamefont{J.}~\bibnamefont{Ye}},
  \bibinfo{author}{\bibfnamefont{D.~W.} \bibnamefont{Vernooy}},
  \bibnamefont{and} \bibinfo{author}{\bibfnamefont{H.~J.}
  \bibnamefont{Kimble}}, \bibinfo{journal}{Physical Review Letters}
  \textbf{\bibinfo{volume}{83}}, \bibinfo{pages}{4987} (\bibinfo{year}{1999}).

\bibitem[{\citenamefont{McKeever et~al.}(2003)\citenamefont{McKeever, Buck,
  Boozer, Kuzmich, N\"{a}gerl, Stamper-Kurn, and Kimble}}]{McKeever2003}
\bibinfo{author}{\bibfnamefont{J.}~\bibnamefont{McKeever}},
  \bibinfo{author}{\bibfnamefont{J.~R.} \bibnamefont{Buck}},
  \bibinfo{author}{\bibfnamefont{A.~D.} \bibnamefont{Boozer}},
  \bibinfo{author}{\bibfnamefont{A.}~\bibnamefont{Kuzmich}},
  \bibinfo{author}{\bibfnamefont{H.-C.} \bibnamefont{N\"{a}gerl}},
  \bibinfo{author}{\bibfnamefont{D.~M.} \bibnamefont{Stamper-Kurn}},
  \bibnamefont{and} \bibinfo{author}{\bibfnamefont{H.~J.}
  \bibnamefont{Kimble}}, \bibinfo{journal}{Physical Review Letters}
  \textbf{\bibinfo{volume}{90}}, \bibinfo{pages}{2} (\bibinfo{year}{2003}).

\bibitem[{\citenamefont{Maunz et~al.}(2004)\citenamefont{Maunz, Puppe,
  Schuster, Syassen, Pinkse, and Rempe}}]{Maunz2004a}
\bibinfo{author}{\bibfnamefont{P.}~\bibnamefont{Maunz}},
  \bibinfo{author}{\bibfnamefont{T.}~\bibnamefont{Puppe}},
  \bibinfo{author}{\bibfnamefont{I.}~\bibnamefont{Schuster}},
  \bibinfo{author}{\bibfnamefont{N.}~\bibnamefont{Syassen}},
  \bibinfo{author}{\bibfnamefont{P.~W.~H.} \bibnamefont{Pinkse}},
  \bibnamefont{and} \bibinfo{author}{\bibfnamefont{G.}~\bibnamefont{Rempe}},
  \bibinfo{journal}{Nature} \textbf{\bibinfo{volume}{428}}, \bibinfo{pages}{50}
  (\bibinfo{year}{2004}).

\bibitem[{\citenamefont{Leibrandt et~al.}(2009)\citenamefont{Leibrandt,
  Labaziewicz, Vuletic, and Chuang}}]{Leibrandt2009}
\bibinfo{author}{\bibfnamefont{D.~R.} \bibnamefont{Leibrandt}},
  \bibinfo{author}{\bibfnamefont{J.}~\bibnamefont{Labaziewicz}},
  \bibinfo{author}{\bibfnamefont{V.}~\bibnamefont{Vuletic}}, \bibnamefont{and}
  \bibinfo{author}{\bibfnamefont{I.~L.} \bibnamefont{Chuang}},
  \bibinfo{journal}{Phys. Rev. Lett.} \textbf{\bibinfo{volume}{103}},
  \bibinfo{pages}{103001} (\bibinfo{year}{2009}).

\bibitem[{\citenamefont{Stute et~al.}(2012)\citenamefont{Stute, Casabone,
  Schindler, Monz, Schmidt, Brandstatter, Northup, and Blatt}}]{Stute2012}
\bibinfo{author}{\bibfnamefont{A.}~\bibnamefont{Stute}},
  \bibinfo{author}{\bibfnamefont{B.}~\bibnamefont{Casabone}},
  \bibinfo{author}{\bibfnamefont{P.}~\bibnamefont{Schindler}},
  \bibinfo{author}{\bibfnamefont{T.}~\bibnamefont{Monz}},
  \bibinfo{author}{\bibfnamefont{P.~O.} \bibnamefont{Schmidt}},
  \bibinfo{author}{\bibfnamefont{B.}~\bibnamefont{Brandstatter}},
  \bibinfo{author}{\bibfnamefont{T.~E.} \bibnamefont{Northup}},
  \bibnamefont{and} \bibinfo{author}{\bibfnamefont{R.}~\bibnamefont{Blatt}},
  \bibinfo{journal}{Nature} \textbf{\bibinfo{volume}{485}},
  \bibinfo{pages}{482} (\bibinfo{year}{2012}).

\bibitem[{\citenamefont{Hechenblaikner
  et~al.}(1998)\citenamefont{Hechenblaikner, Gangl, Horak, and
  Ritsch}}]{Hechenblaikner1998}
\bibinfo{author}{\bibfnamefont{G.}~\bibnamefont{Hechenblaikner}},
  \bibinfo{author}{\bibfnamefont{M.}~\bibnamefont{Gangl}},
  \bibinfo{author}{\bibfnamefont{P.}~\bibnamefont{Horak}}, \bibnamefont{and}
  \bibinfo{author}{\bibfnamefont{H.}~\bibnamefont{Ritsch}},
  \bibinfo{journal}{Physical Review A} \textbf{\bibinfo{volume}{58}},
  \bibinfo{pages}{3030} (\bibinfo{year}{1998}).

\bibitem[{\citenamefont{Gangl and Ritsch}(2000)}]{Gangl2000}
\bibinfo{author}{\bibfnamefont{M.}~\bibnamefont{Gangl}} \bibnamefont{and}
  \bibinfo{author}{\bibfnamefont{H.}~\bibnamefont{Ritsch}},
  \bibinfo{journal}{European Physical Journal D} \textbf{\bibinfo{volume}{8}},
  \bibinfo{pages}{29} (\bibinfo{year}{2000}).

\bibitem[{\citenamefont{Favero et~al.}(2009)\citenamefont{Favero, Stapfner,
  Hunger, Paulitschke, Reichel, Lorenz, Weig, and Karrai}}]{Favero2009b}
\bibinfo{author}{\bibfnamefont{I.}~\bibnamefont{Favero}},
  \bibinfo{author}{\bibfnamefont{S.}~\bibnamefont{Stapfner}},
  \bibinfo{author}{\bibfnamefont{D.}~\bibnamefont{Hunger}},
  \bibinfo{author}{\bibfnamefont{P.}~\bibnamefont{Paulitschke}},
  \bibinfo{author}{\bibfnamefont{J.}~\bibnamefont{Reichel}},
  \bibinfo{author}{\bibfnamefont{H.}~\bibnamefont{Lorenz}},
  \bibinfo{author}{\bibfnamefont{E.~M.} \bibnamefont{Weig}}, \bibnamefont{and}
  \bibinfo{author}{\bibfnamefont{K.}~\bibnamefont{Karrai}},
  \bibinfo{journal}{Optics Express} \textbf{\bibinfo{volume}{17}},
  \bibinfo{pages}{16} (\bibinfo{year}{2009}).

\bibitem[{\citenamefont{Anetsberger et~al.}(2009)\citenamefont{Anetsberger,
  Arcizet, Unterreithmeier, Rivi\`{e}re, Schliesser, Weig, Kotthaus, P., and
  Kippenberg}}]{Anetsberger2009a}
\bibinfo{author}{\bibfnamefont{G.}~\bibnamefont{Anetsberger}},
  \bibinfo{author}{\bibfnamefont{O.}~\bibnamefont{Arcizet}},
  \bibinfo{author}{\bibfnamefont{Q.~P.} \bibnamefont{Unterreithmeier}},
  \bibinfo{author}{\bibfnamefont{R.}~\bibnamefont{Rivi\`{e}re}},
  \bibinfo{author}{\bibfnamefont{A.}~\bibnamefont{Schliesser}},
  \bibinfo{author}{\bibfnamefont{E.~M.} \bibnamefont{Weig}},
  \bibinfo{author}{\bibnamefont{Kotthaus}},
  \bibinfo{author}{\bibfnamefont{J.}~\bibnamefont{P.}}, \bibnamefont{and}
  \bibinfo{author}{\bibfnamefont{T.~J.} \bibnamefont{Kippenberg}},
  \bibinfo{journal}{Nature Physics} \textbf{\bibinfo{volume}{5}},
  \bibinfo{pages}{909} (\bibinfo{year}{2009}).

\bibitem[{\citenamefont{Chan et~al.}(2011)\citenamefont{Chan, Alegre,
  Safavi-Naeini, Hill, Krause, Gr\"{o}blacher, Aspelmeyer, and
  Painter}}]{Chan2011b}
\bibinfo{author}{\bibfnamefont{J.}~\bibnamefont{Chan}},
  \bibinfo{author}{\bibfnamefont{T.~P.~M.} \bibnamefont{Alegre}},
  \bibinfo{author}{\bibfnamefont{A.~H.} \bibnamefont{Safavi-Naeini}},
  \bibinfo{author}{\bibfnamefont{J.~T.} \bibnamefont{Hill}},
  \bibinfo{author}{\bibfnamefont{A.}~\bibnamefont{Krause}},
  \bibinfo{author}{\bibfnamefont{S.}~\bibnamefont{Gr\"{o}blacher}},
  \bibinfo{author}{\bibfnamefont{M.}~\bibnamefont{Aspelmeyer}},
  \bibnamefont{and} \bibinfo{author}{\bibfnamefont{O.}~\bibnamefont{Painter}},
  \bibinfo{journal}{Nature} \textbf{\bibinfo{volume}{478}}, \bibinfo{pages}{89}
  (\bibinfo{year}{2011}).

\bibitem[{\citenamefont{Thompson et~al.}(2008)\citenamefont{Thompson, Zwickl,
  Jayich, Marquardt, Girvin, and Harris}}]{Thompson_etal2008}
\bibinfo{author}{\bibfnamefont{J.~D.} \bibnamefont{Thompson}},
  \bibinfo{author}{\bibfnamefont{B.~M.} \bibnamefont{Zwickl}},
  \bibinfo{author}{\bibfnamefont{A.~M.} \bibnamefont{Jayich}},
  \bibinfo{author}{\bibfnamefont{F.}~\bibnamefont{Marquardt}},
  \bibinfo{author}{\bibfnamefont{S.~M.} \bibnamefont{Girvin}},
  \bibnamefont{and} \bibinfo{author}{\bibfnamefont{J.~G.~E.}
  \bibnamefont{Harris}}, \bibinfo{journal}{Nature}
  \textbf{\bibinfo{volume}{452}}, \bibinfo{pages}{72} (\bibinfo{year}{2008}).

\bibitem[{\citenamefont{Teufel et~al.}(2011)\citenamefont{Teufel, Donner, Li,
  Harlow, Allman, Cicak, Sirois, Whittaker, Lehnert, and
  Simmonds}}]{Teufel2011}
\bibinfo{author}{\bibfnamefont{J.~D.} \bibnamefont{Teufel}},
  \bibinfo{author}{\bibfnamefont{T.}~\bibnamefont{Donner}},
  \bibinfo{author}{\bibfnamefont{D.}~\bibnamefont{Li}},
  \bibinfo{author}{\bibfnamefont{J.~W.} \bibnamefont{Harlow}},
  \bibinfo{author}{\bibfnamefont{M.~S.} \bibnamefont{Allman}},
  \bibinfo{author}{\bibfnamefont{K.}~\bibnamefont{Cicak}},
  \bibinfo{author}{\bibfnamefont{A.~J.} \bibnamefont{Sirois}},
  \bibinfo{author}{\bibfnamefont{J.~D.} \bibnamefont{Whittaker}},
  \bibinfo{author}{\bibfnamefont{K.~W.} \bibnamefont{Lehnert}},
  \bibnamefont{and} \bibinfo{author}{\bibfnamefont{R.~W.}
  \bibnamefont{Simmonds}}, \bibinfo{journal}{Nature}
  \textbf{\bibinfo{volume}{475}}, \bibinfo{pages}{359} (\bibinfo{year}{2011}).

\bibitem[{\citenamefont{Purdy et~al.}(2013)\citenamefont{Purdy, Peterson, and
  Regal}}]{Purdy2013}
\bibinfo{author}{\bibfnamefont{T.~P.} \bibnamefont{Purdy}},
  \bibinfo{author}{\bibfnamefont{R.~W.} \bibnamefont{Peterson}},
  \bibnamefont{and} \bibinfo{author}{\bibfnamefont{C.~A.} \bibnamefont{Regal}},
  \bibinfo{journal}{Science} \textbf{\bibinfo{volume}{339}},
  \bibinfo{pages}{801} (\bibinfo{year}{2013}).

\bibitem[{\citenamefont{Safavi-Naeini et~al.}(2012)\citenamefont{Safavi-Naeini,
  Chan, Hill, Mayer~Alegre, Krause, and Painter}}]{Safavi-Naeini2012}
\bibinfo{author}{\bibfnamefont{A.~H.} \bibnamefont{Safavi-Naeini}},
  \bibinfo{author}{\bibfnamefont{J.}~\bibnamefont{Chan}},
  \bibinfo{author}{\bibfnamefont{J.~T.} \bibnamefont{Hill}},
  \bibinfo{author}{\bibfnamefont{T.~P.} \bibnamefont{Mayer~Alegre}},
  \bibinfo{author}{\bibfnamefont{A.}~\bibnamefont{Krause}}, \bibnamefont{and}
  \bibinfo{author}{\bibfnamefont{O.}~\bibnamefont{Painter}},
  \bibinfo{journal}{Phys. Rev. Lett.} \textbf{\bibinfo{volume}{108}},
  \bibinfo{pages}{033602} (\bibinfo{year}{2012}).

\bibitem[{\citenamefont{Murch et~al.}(2008)\citenamefont{Murch, Moore, Gupta,
  and Stamper-Kurn}}]{Murch2008}
\bibinfo{author}{\bibfnamefont{K.~W.} \bibnamefont{Murch}},
  \bibinfo{author}{\bibfnamefont{K.~L.} \bibnamefont{Moore}},
  \bibinfo{author}{\bibfnamefont{S.}~\bibnamefont{Gupta}}, \bibnamefont{and}
  \bibinfo{author}{\bibfnamefont{D.~M.} \bibnamefont{Stamper-Kurn}},
  \bibinfo{journal}{Nature Physics} \textbf{\bibinfo{volume}{4}},
  \bibinfo{pages}{561} (\bibinfo{year}{2008}).

\bibitem[{\citenamefont{Purdy et~al.}(2010)\citenamefont{Purdy, Brooks, Botter,
  Brahms, Ma, and Stamper-Kurn}}]{Purdy2010}
\bibinfo{author}{\bibfnamefont{T.}~\bibnamefont{Purdy}},
  \bibinfo{author}{\bibfnamefont{D.}~\bibnamefont{Brooks}},
  \bibinfo{author}{\bibfnamefont{T.}~\bibnamefont{Botter}},
  \bibinfo{author}{\bibfnamefont{N.}~\bibnamefont{Brahms}},
  \bibinfo{author}{\bibfnamefont{Z.-Y.} \bibnamefont{Ma}}, \bibnamefont{and}
  \bibinfo{author}{\bibfnamefont{D.~M.} \bibnamefont{Stamper-Kurn}},
  \bibinfo{journal}{Phys. Rev. Lett.} \textbf{\bibinfo{volume}{105}},
  \bibinfo{pages}{133602} (\bibinfo{year}{2010}).

\bibitem[{\citenamefont{Schleier-Smith
  et~al.}(2011)\citenamefont{Schleier-Smith, Leroux, Zhang, {Van Camp}, and
  Vuleti\'{c}}}]{Schleier-smith2011a}
\bibinfo{author}{\bibfnamefont{M.~H.} \bibnamefont{Schleier-Smith}},
  \bibinfo{author}{\bibfnamefont{I.~D.} \bibnamefont{Leroux}},
  \bibinfo{author}{\bibfnamefont{H.}~\bibnamefont{Zhang}},
  \bibinfo{author}{\bibfnamefont{M.~A.} \bibnamefont{{Van Camp}}},
  \bibnamefont{and}
  \bibinfo{author}{\bibfnamefont{V.}~\bibnamefont{Vuleti\'{c}}},
  \bibinfo{journal}{Physical Review Letters} \textbf{\bibinfo{volume}{107}},
  \bibinfo{pages}{143005} (\bibinfo{year}{2011}).

\bibitem[{\citenamefont{Kaltenbaek et~al.}(2012)\citenamefont{Kaltenbaek,
  Hechenblaikner, Kiesel, Romero-Isart, Schwab, Johann, and
  Aspelmeyer}}]{Kaltenbaek2012a}
\bibinfo{author}{\bibfnamefont{R.}~\bibnamefont{Kaltenbaek}},
  \bibinfo{author}{\bibfnamefont{G.}~\bibnamefont{Hechenblaikner}},
  \bibinfo{author}{\bibfnamefont{N.}~\bibnamefont{Kiesel}},
  \bibinfo{author}{\bibfnamefont{O.}~\bibnamefont{Romero-Isart}},
  \bibinfo{author}{\bibfnamefont{K.}~\bibnamefont{Schwab}},
  \bibinfo{author}{\bibfnamefont{U.}~\bibnamefont{Johann}}, \bibnamefont{and}
  \bibinfo{author}{\bibfnamefont{M.}~\bibnamefont{Aspelmeyer}},
  \bibinfo{journal}{Experimental Astronomy} \textbf{\bibinfo{volume}{34}},
  \bibinfo{pages}{123} (\bibinfo{year}{2012}).

\bibitem[{\citenamefont{Ashkin}(2007)}]{Ashkin2007}
\bibinfo{author}{\bibfnamefont{A.}~\bibnamefont{Ashkin}},
  \emph{\bibinfo{title}{Optical trapping and manipulation of neutral particles
  using lasers}} (\bibinfo{publisher}{World Scientific Pub Co},
  \bibinfo{year}{2007}).

\bibitem[{\citenamefont{Nimmrichter et~al.}(2010)\citenamefont{Nimmrichter,
  Hammerer, Asenbaum, Ritsch, and Arndt}}]{Nimmrichter2010b}
\bibinfo{author}{\bibfnamefont{S.}~\bibnamefont{Nimmrichter}},
  \bibinfo{author}{\bibfnamefont{K.}~\bibnamefont{Hammerer}},
  \bibinfo{author}{\bibfnamefont{P.}~\bibnamefont{Asenbaum}},
  \bibinfo{author}{\bibfnamefont{H.}~\bibnamefont{Ritsch}}, \bibnamefont{and}
  \bibinfo{author}{\bibfnamefont{M.}~\bibnamefont{Arndt}},
  \bibinfo{journal}{New Journal of Physics} \textbf{\bibinfo{volume}{12}},
  \bibinfo{pages}{083003} (\bibinfo{year}{2010}).

\bibitem[{\citenamefont{Romero-Isart
  et~al.}(2011{\natexlab{b}})\citenamefont{Romero-Isart, Pflanzer, Juan,
  Quidant, Kiesel, Aspelmeyer, and Cirac}}]{Romero-Isart2011}
\bibinfo{author}{\bibfnamefont{O.}~\bibnamefont{Romero-Isart}},
  \bibinfo{author}{\bibfnamefont{A.}~\bibnamefont{Pflanzer}},
  \bibinfo{author}{\bibfnamefont{M.}~\bibnamefont{Juan}},
  \bibinfo{author}{\bibfnamefont{R.}~\bibnamefont{Quidant}},
  \bibinfo{author}{\bibfnamefont{N.}~\bibnamefont{Kiesel}},
  \bibinfo{author}{\bibfnamefont{M.}~\bibnamefont{Aspelmeyer}},
  \bibnamefont{and} \bibinfo{author}{\bibfnamefont{J.~I.} \bibnamefont{Cirac}},
  \bibinfo{journal}{Phys. Rev. A} \textbf{\bibinfo{volume}{83}},
  \bibinfo{pages}{013803} (\bibinfo{year}{2011}{\natexlab{b}}).

\bibitem[{\citenamefont{Barker and Shneider}(2010)}]{Barker2010a}
\bibinfo{author}{\bibfnamefont{P.~F.} \bibnamefont{Barker}} \bibnamefont{and}
  \bibinfo{author}{\bibfnamefont{M.~N.} \bibnamefont{Shneider}},
  \bibinfo{journal}{Phys. Rev. A} \textbf{\bibinfo{volume}{81}},
  \bibinfo{pages}{023826} (\bibinfo{year}{2010}).

\bibitem[{\citenamefont{Stamper-Kurn}(2012)}]{Stamper-Kurn2012}
\bibinfo{author}{\bibfnamefont{D.}~\bibnamefont{Stamper-Kurn}},
  \bibinfo{journal}{arXiv:1204.4351v1}  (\bibinfo{year}{2012}).

\bibitem[{\citenamefont{Pender et~al.}(2012)\citenamefont{Pender, Barker,
  Marquardt, Millen, and Monteiro}}]{Pender2012}
\bibinfo{author}{\bibfnamefont{G.~A.~T.} \bibnamefont{Pender}},
  \bibinfo{author}{\bibfnamefont{P.~F.} \bibnamefont{Barker}},
  \bibinfo{author}{\bibfnamefont{F.}~\bibnamefont{Marquardt}},
  \bibinfo{author}{\bibfnamefont{J.}~\bibnamefont{Millen}}, \bibnamefont{and}
  \bibinfo{author}{\bibfnamefont{T.~S.} \bibnamefont{Monteiro}},
  \bibinfo{journal}{Phys. Rev. A} \textbf{\bibinfo{volume}{85}},
  \bibinfo{pages}{021802} (\bibinfo{year}{2012}).

\bibitem[{\citenamefont{Monteiro et~al.}(2013)\citenamefont{Monteiro, Millen,
  Pender, Marquardt, Chang, and Barker}}]{Monteiro2013}
\bibinfo{author}{\bibfnamefont{T.~S.} \bibnamefont{Monteiro}},
  \bibinfo{author}{\bibfnamefont{J.}~\bibnamefont{Millen}},
  \bibinfo{author}{\bibfnamefont{G.~A.~T.} \bibnamefont{Pender}},
  \bibinfo{author}{\bibfnamefont{F.}~\bibnamefont{Marquardt}},
  \bibinfo{author}{\bibfnamefont{D.}~\bibnamefont{Chang}}, \bibnamefont{and}
  \bibinfo{author}{\bibfnamefont{P.~F.} \bibnamefont{Barker}},
  \bibinfo{journal}{New Journal of Physics} \textbf{\bibinfo{volume}{15}},
  \bibinfo{pages}{015001} (\bibinfo{year}{2013}).

\bibitem[{\citenamefont{Paternostro et~al.}(2006)\citenamefont{Paternostro,
  Gigan, Kim, Blaser, B\"{o}hm, and Aspelmeyer}}]{Paternostro2006}
\bibinfo{author}{\bibfnamefont{M.}~\bibnamefont{Paternostro}},
  \bibinfo{author}{\bibfnamefont{S.}~\bibnamefont{Gigan}},
  \bibinfo{author}{\bibfnamefont{M.~S.} \bibnamefont{Kim}},
  \bibinfo{author}{\bibfnamefont{F.}~\bibnamefont{Blaser}},
  \bibinfo{author}{\bibfnamefont{H.~R.} \bibnamefont{B\"{o}hm}},
  \bibnamefont{and}
  \bibinfo{author}{\bibfnamefont{M.}~\bibnamefont{Aspelmeyer}},
  \bibinfo{journal}{New Journal of Physics} \textbf{\bibinfo{volume}{8}},
  \bibinfo{pages}{107} (\bibinfo{year}{2006}).

\bibitem[{\citenamefont{Ashkin and Dziedzic}(1976)}]{Ashkin1976a}
\bibinfo{author}{\bibfnamefont{A.}~\bibnamefont{Ashkin}} \bibnamefont{and}
  \bibinfo{author}{\bibfnamefont{J.~M.} \bibnamefont{Dziedzic}},
  \bibinfo{journal}{Applied Physics Letters} \textbf{\bibinfo{volume}{28}},
  \bibinfo{pages}{333} (\bibinfo{year}{1976}).

\bibitem[{\citenamefont{Akram et~al.}(2010)\citenamefont{Akram, Kiesel,
  Aspelmeyer, and Milburn}}]{Akram2010}
\bibinfo{author}{\bibfnamefont{U.}~\bibnamefont{Akram}},
  \bibinfo{author}{\bibfnamefont{N.}~\bibnamefont{Kiesel}},
  \bibinfo{author}{\bibfnamefont{M.}~\bibnamefont{Aspelmeyer}},
  \bibnamefont{and} \bibinfo{author}{\bibfnamefont{G.~J.}
  \bibnamefont{Milburn}}, \bibinfo{journal}{New Journal of Physics}
  \textbf{\bibinfo{volume}{12}}, \bibinfo{pages}{083030}
  (\bibinfo{year}{2010}).

\bibitem[{\citenamefont{Romero-Isart}(2011)}]{Romero-Isart2011b}
\bibinfo{author}{\bibfnamefont{O.}~\bibnamefont{Romero-Isart}},
  \bibinfo{journal}{Phys. Rev. A} \textbf{\bibinfo{volume}{84}},
  \bibinfo{pages}{052121} (\bibinfo{year}{2011}).

\bibitem[{\citenamefont{Lechner et~al.}(2013)\citenamefont{Lechner, Habraken,
  Kiesel, Aspelmeyer, and Zoller}}]{Lechner2012}
\bibinfo{author}{\bibfnamefont{W.}~\bibnamefont{Lechner}},
  \bibinfo{author}{\bibfnamefont{S.~J.~M.} \bibnamefont{Habraken}},
  \bibinfo{author}{\bibfnamefont{N.}~\bibnamefont{Kiesel}},
  \bibinfo{author}{\bibfnamefont{M.}~\bibnamefont{Aspelmeyer}},
  \bibnamefont{and} \bibinfo{author}{\bibfnamefont{P.}~\bibnamefont{Zoller}},
  \bibinfo{journal}{Phys. Rev. Lett.} \textbf{\bibinfo{volume}{110}},
  \bibinfo{pages}{143604} (\bibinfo{year}{2013}).

\bibitem[{\citenamefont{Summers et~al.}(2008)\citenamefont{Summers, Burnham,
  and McGloin}}]{Summers2008}
\bibinfo{author}{\bibfnamefont{M.~D.} \bibnamefont{Summers}},
  \bibinfo{author}{\bibfnamefont{D.~R.} \bibnamefont{Burnham}},
  \bibnamefont{and} \bibinfo{author}{\bibfnamefont{D.}~\bibnamefont{McGloin}},
  \bibinfo{journal}{Optics Express} \textbf{\bibinfo{volume}{16}},
  \bibinfo{pages}{7739} (\bibinfo{year}{2008}).

\bibitem[{\citenamefont{Saleh and Teich}(2007)}]{saleh2007}
\bibinfo{author}{\bibfnamefont{B.~E.~A.} \bibnamefont{Saleh}} \bibnamefont{and}
  \bibinfo{author}{\bibfnamefont{M.~C.} \bibnamefont{Teich}},
  \emph{\bibinfo{title}{Fundamentals of photonics}}
  (\bibinfo{publisher}{Wiley-Interscience}, \bibinfo{year}{2007}).

\bibitem[{\citenamefont{Genes et~al.}(2008)\citenamefont{Genes, Vitali,
  Tombesi, Gigan, and Aspelmeyer}}]{Genes2008a}
\bibinfo{author}{\bibfnamefont{C.}~\bibnamefont{Genes}},
  \bibinfo{author}{\bibfnamefont{D.}~\bibnamefont{Vitali}},
  \bibinfo{author}{\bibfnamefont{P.}~\bibnamefont{Tombesi}},
  \bibinfo{author}{\bibfnamefont{S.}~\bibnamefont{Gigan}}, \bibnamefont{and}
  \bibinfo{author}{\bibfnamefont{M.}~\bibnamefont{Aspelmeyer}},
  \bibinfo{journal}{Physical Review A} \textbf{\bibinfo{volume}{77}},
  \bibinfo{pages}{33804} (\bibinfo{year}{2008}).

\bibitem[{\citenamefont{Gardiner and Zoller}(2004)}]{gardiner2004}
\bibinfo{author}{\bibfnamefont{C.}~\bibnamefont{Gardiner}} \bibnamefont{and}
  \bibinfo{author}{\bibfnamefont{P.}~\bibnamefont{Zoller}},
  \emph{\bibinfo{title}{Quantum noise: a handbook of Markovian and
  non-Markovian quantum stochastic methods with applications to quantum
  optics}}, vol.~\bibinfo{volume}{56} (\bibinfo{publisher}{Springer},
  \bibinfo{year}{2004}).

\bibitem[{\citenamefont{Bachor and Ralph}(2004)}]{bachor2004guide}
\bibinfo{author}{\bibfnamefont{H.-A.} \bibnamefont{Bachor}} \bibnamefont{and}
  \bibinfo{author}{\bibfnamefont{T.~C.} \bibnamefont{Ralph}},
  \emph{\bibinfo{title}{A Guide to Experiments in Quantum Optics}}
  (\bibinfo{publisher}{Wiley-VCH}, \bibinfo{year}{2004}).

\bibitem[{\citenamefont{Beresnev et~al.}(1990)\citenamefont{Beresnev, Chernyak,
  and Fomyagin}}]{Beresnev1990}
\bibinfo{author}{\bibfnamefont{S.~A.} \bibnamefont{Beresnev}},
  \bibinfo{author}{\bibfnamefont{V.~G.} \bibnamefont{Chernyak}},
  \bibnamefont{and} \bibinfo{author}{\bibfnamefont{G.~A.}
  \bibnamefont{Fomyagin}}, \bibinfo{journal}{Journal of Fluid Mechanics}
  \textbf{\bibinfo{volume}{219}}, \bibinfo{pages}{405} (\bibinfo{year}{1990}).

\end{thebibliography}

\end{document}